\documentclass[useAMS,usenatbib,fleqn]{mnras}
\usepackage[english]{babel}
\usepackage{amsmath}
\usepackage{fancyhdr}
\usepackage{amssymb}
\usepackage{amsfonts}
\usepackage{graphicx}
\usepackage{color}
\usepackage{natbib}
\usepackage{multicol}
\usepackage{layout}
\usepackage{times}

\input{psfig}
\def\beq{\begin{equation}}
\def\eeq{\end{equation}}
\def\barray{\begin{eqnarray}}
\def\earray{\end{eqnarray}}
\def\beqarray{\begin{eqnarray}}
\def\eeqarray{\end{eqnarray}}

\def\rmc{{\rm c}}
\def\rmd{{\rm d}}

\def\rmh{{\rm h}}

\def\rms{{\rm s}}

\def\rmx{{\rm x}}
\def\rmy{{\rm y}}

\def\rmM{{\rm M}}

\def\calH{{\cal H}}

\newcommand{\Msun}{\>{\rm M_{\odot}}}

\def\gtsima{$\; \buildrel > \over \sim \;$}
\def\ltsima{$\; \buildrel < \over \sim \;$}
\def\prosima{$\; \buildrel \propto \over \sim \;$}
\def\gsim{\lower.7ex\hbox{\gtsima}}
\def\lsim{\lower.7ex\hbox{\ltsima}}
\def\simgt{\lower.7ex\hbox{\gtsima}}
\def\simlt{\lower.7ex\hbox{\ltsima}}
\def\simpr{\lower.7ex\hbox{\prosima}}
\def\la{\lsim}
\def\ga{\gsim}

\newdimen\hssize
\newdimen\hdsize
\title{The stellar-to-halo mass relation of GAMA galaxies from 100 square degrees of KiDS weak lensing data}
\author[Edo van Uitert et al.]
{Edo van Uitert $^{1,2}$\thanks{vuitert@ucl.ac.uk}, Marcello Cacciato$^3$, Henk Hoekstra$^3$, Margot Brouwer$^3$, Crist\'obal Sif\'on$^3$, \and Massimo Viola$^3$, Ivan Baldry$^4$, Joss Bland-Hawthorn$^5$, Sarah Brough$^6$, M. J. I. Brown$^7$, \and  Ami Choi$^8$, Simon P. Driver$^{9,10}$, Thomas Erben$^2$, Catherine Heymans$^8$, Hendrik Hildebrandt$^2$, \and Benjamin Joachimi $^1$, Konrad Kuijken$^3$, Jochen Liske$^{11}$, Jon Loveday$^{12}$, John McFarland$^{13}$, \and Lance Miller$^{14}$, Reiko Nakajima$^2$, John Peacock$^8$, Mario Radovich$^{15}$, A. S. G. Robotham$^{16}$, \and Peter Schneider$^2$, Gert Sikkema$^{13}$, Edward N. Taylor$^{17}$, Gijs Verdoes Kleijn$^{13}$ \\ 
\\
$^1$ University College London, Gower Street, London WC1E 6BT, UK \\ 
$^2$ Argelander-Institut f\"ur Astronomie, Auf dem H\"ugel 71, 53121 Bonn, Germany \\ 
$^3$ Leiden Observatory, Leiden University, Niels Bohrweg 2, NL-2333 CA Leiden, The Netherlands \\
$^4$ Astrophysics Research Institute, Liverpool John Moores University, IC2, Liverpool Science Park, 146 Brownlow Hill, Liverpool L3 5RF, UK \\
$^5$ Sydney Institute for Astronomy, School of Physics A28, University of Sydney, NSW 2006, Australia \\
$^6$ Australian Astronomical Observatory, PO Box 915, North Ryde, NSW 1670, Australia \\
$^7$ School of Physics and Astronomy, Monash University, Clayton, Victoria 3800, Australia \\
$^8$ Scottish Universities Physics Alliance, Institute for Astronomy, University of Edinburgh, Royal Observatory, Blackford Hill, Edinburgh EH9 3HJ, UK  \\
$^9$ International Centre for Radio Astronomy Research (ICRAR), The University of Western Australia, 35 Stirling Highway, Crawley, WA 6009, Australia \\
$^{10}$ Scottish Universities Physics Alliance, School of Physics \& Astronomy, University of St Andrews, North Haugh, St Andrews KY16 9SS, UK \\
$^{11}$ Hamburger Sternwarte, Universit{\"a}t Hamburg, Gojenbergsweg 112, 21029 Hamburg, Germany \\
$^{12}$ Astronomy Centre, Department of Physics and Astronomy, University of Sussex, Falmer, Brighton BN1 9QH, UK \\
$^{13}$ Kapteyn Astronomical Institute, University of Groningen, PO Box 800, NL-9700 AV Groningen, the Netherlands \\
$^{14}$ Department of Physics, Oxford University, Keble Road, Oxford OX1 3RH, UK \\
$^{15}$ INAF - Osservatorio Astronomico di Padova, via dell`Osservatorio 5, I-35122 Padova, Italy \\
$^{16}$ International Centre for Radio Astronomy Research, University of Western Australia, 35 Stirling Highway, Crawley, WA 6009, Australia  \\
$^{17}$ School of Physics, The University of Melbourne, Parkville, VIC 3010, Australia  \\
}

\pubyear{2016}

\begin{document}
\maketitle

\begin{abstract}
We study the stellar-to-halo mass relation of central galaxies in the range \mbox{$9.7<\log_{10}(M_*/h^{-2}M_\odot)<11.7$} and $z<0.4$, obtained from a combined analysis of the Kilo Degree Survey (KiDS) and the Galaxy And Mass Assembly (GAMA) survey. We use \mbox{$\sim$100 deg$^2$} of KiDS data to study the lensing signal around galaxies for which spectroscopic redshifts and stellar masses were determined by GAMA. We show that lensing alone results in poor constraints on the stellar-to-halo mass relation due to a degeneracy between the satellite fraction and the halo mass, which is lifted when we simultaneously fit the stellar mass function. At $M_*>5\times10^{10}h^{-2}M_\odot$, the stellar mass increases with halo mass as $\sim$$M_\rmh^{0.25}$. The ratio of dark matter to stellar mass has a minimum at a halo mass of $8\times10^{11}h^{-1}M_\odot$ with a value of $M_\rmh/M_*=56_{-10}^{+16}$ [$h$]. We also use the GAMA group catalogue to select centrals and satellites in groups with five or more members, which trace regions in space where the local matter density is higher than average, and determine for the first time the stellar-to-halo mass relation in these denser environments. We find no significant differences compared to the relation from the full sample, which suggests that the stellar-to-halo mass relation does not vary strongly with local density. Furthermore, we find that the stellar-to-halo mass relation of central galaxies can also be obtained by modelling the lensing signal and stellar mass function of satellite galaxies only, which shows that the assumptions to model the satellite contribution in the halo model do not significantly bias the stellar-to-halo mass relation. Finally, we show that the combination of weak lensing with the stellar mass function can be used to test the purity of group catalogues.
\end{abstract}


\section{Introduction}  
Galaxies form and evolve in dark matter haloes. Larger haloes attract on average more baryons and host larger, more massive galaxies. The exact relation between the baryonic properties of galaxies and their dark matter haloes is complex, however, as various astrophysical processes are involved. These include supernova and AGN feedback \citep[see e.g.][]{Benson10}, whose relative importances generally depend on halo mass in a way that is not accurately known, but environmental effects also play an important role. Measuring projections of these relations, such as the stellar-to-halo mass relation, helps us to gain insight into these processes and their mass dependences, and provides valuable references for comparisons for numerical simulations that model galaxy formation and evolution \citep[e.g.][]{Munshi13,Kannan14}.  \\
\indent The stellar-to-halo mass relation has been studied with a variety of methods, including indirect techniques such as abundance matching \citep[e.g.][]{Behroozi10,Moster13} or galaxy clustering \citep[e.g.][]{Wake11,Guo14}, which can only be interpreted within a cosmological framework (e.g. $\Lambda$CDM). Satellite kinematics offer a direct way to measure halo mass \citep[e.g.][]{Norberg08,Wojtak13}, but this approach is relatively expensive as it requires spectroscopy for large samples of satellites. Weak gravitational lensing offers another powerful method that enables average halo mass measurements for ensembles of galaxies \citep[e.g.][]{Mandelbaum06,Velander14}. Recently, various groups have combined different probes \citep[e.g.][]{Leauthaud12,Coupon15}, which enable more stringent constraints on the stellar-to-halo mass relation by breaking degeneracies between model parameters. A coherent picture is emerging from these studies: the stellar-to-halo mass relation of central galaxies can be described by a double power law, with a transition at a pivot mass where the accumulated star formation has been most efficient. At higher masses, AGN feedback is thought to suppress star formation, whilst at lower masses, supernova feedback suppresses it. This pivot mass coincides with the location where the stellar mass growth in galaxies turns from being in-situ dominated to merger dominated \citep{Robotham14}.   \\
\indent  Most galaxies can be roughly divided into two classes, i.e. red, `early-types' whose star formation has been quenched, and blue, `late-types' that are actively forming stars. These are also crudely related to different environments and morphologies. The differences in their appearances point at different formation histories. Their stellar-to-halo mass relations may contain information of the underlying physical processes that caused these differences. Hence it is natural to measure the stellar-to-halo mass relations of red and blue galaxies separately \citep[e.g.][]{Mandelbaum06,VanUitert11,More11,Velander14,Wojtak13,Tinker13,Hudson15}. The main result of the aforementioned studies is that at stellar masses below $\sim10^{11}M_\odot$, red and blue galaxies that are centrals (i.e. not a satellite of a larger system) reside in haloes with comparable masses. At higher stellar masses, the halo masses of red galaxies are larger at low redshift, but smaller at high redshift at a given stellar mass. \citet{Tinker13} interpret this similarity in halo mass at the low-mass end as evidence that these red galaxies have only recently been quenched; the difference at the high-mass end is interpreted as evidence that blue galaxies have a relatively larger stellar mass growth in recent times, compared to red galaxies. \\
\indent As red galaxies preferentially reside in dense environments such as galaxy groups and clusters, it appears that local density is the main driver behind the variation of the stellar-to-halo mass relation, and that the change in colour is simply a consequence of quenching, as was already hypothesized in \citet{Mandelbaum06}. This scenario could be verified by measuring the stellar-to-halo mass relation for galaxies in different environments. This requires a galaxy catalogue including stellar masses and environmental information, plus a method to measure masses. Weak gravitational lensing offers a particularly attractive way of measuring average halo masses of samples of galaxies, as it measures the total projected matter density along the line of sight, without any assumption about the physical state of the matter, out to scales that are inaccessible to other gravitational probes. \\
\indent These conditions are provided by combining two surveys: the Kilo Degree Survey (KiDS) and the Galaxy And Mass Assembly (GAMA) survey. GAMA is a spectroscopic survey for galaxies with $r<19.8$ that is highly complete \citep{Driver09,Driver11,Liske15}, facilitating the construction of a reliable group catalogue \citep{Robotham11}. GAMA is completely covered by the KiDS survey \citep{DeJong13}, an ongoing weak lensing survey which will eventually cover 1500 deg$^2$ of sky in the $ugri$-bands. In this work, we study the lensing signal around $\sim$100\,000 GAMA galaxies using sources from the $\sim$100 deg$^2$ of KiDS imaging data overlapping with the GAMA survey from the first and second publicly available KiDS-DR1/2 data release \citep{Kuijken15}. \\
\indent The outline of this paper is as follows. In Sect. \ref{sec_data} we describe the data reduction and lensing analysis, and introduce the halo model that we fit to our data. The stellar-to-halo mass relation of the full sample is presented in Sect. \ref{sec_msmh}. In Sect. \ref{sec_envir}, we measure this relation for centrals and satellites in groups with a multiplicity $N_{\rm fof}\geq5$. We conclude in Sect. \ref{sec_conc}. Throughout the paper we assume a Planck cosmology \citep{Planck14} with \mbox{$\sigma_8=0.829$}, $\Omega_{\Lambda}=0.685$, $\Omega_{\rm M}=0.315$, \mbox{$\Omega_{\rm b}h^2=0.02205$} and $n_s=0.9603$. Halo masses are defined as \mbox{$M_{\rm h}\equiv 4 \pi(200 {\bar \rho_{\rm m}})R_{200}^3/3$}, with $R_{200}$ the radius of a sphere that encompasses an average density of 200 times the comoving matter density, $\bar{\rho}_{\rm m}=8.74\cdot10^{10} h^2 M_\odot/{\rm Mpc}^3$, at the redshift of the lens. All distances quoted are in comoving (rather than physical) units unless explicitly stated otherwise.

\section{Analysis}\label{sec_data}
\subsection{KiDS}\label{sec_data_KiDS}
To study the weak-lensing signal around galaxies, we use the shape and photometric redshift catalogues from the Kilo Degree Survey \citep[KiDS;][]{Kuijken15}. KiDS is a large optical imaging survey which will cover 1500 deg$^2$ in $u$, $g$, $r$ and $i$ to magnitude limits of 24.2, 25.1, 24.9 and 23.7 (5$\sigma$ in a $2''$ aperture), respectively. Photometry in 5 infrared bands of the same area will become available from the VISTA Kilo-degree Infrared Galaxy (VIKING) survey \citep{Edge13}. The optical observations are carried out with the VLT Survey Telescope (VST) using the 1 deg$^2$ imager OmegaCAM, which consists of 32 CCDs of 2048$\times$4096 pixels each and has a pixel size of 0.214$''$. In this paper, weak lensing results are based on observations of 109 KiDS tiles\footnote{A tile is an observation of a pointing on the sky} that overlap with the GAMA survey, and have been covered in all four optical bands and released to ESO as part of the first and second KiDS-DR1/2 data releases. The effective area after accounting for masks and overlaps between tiles is 75.1 square degrees. The image reduction and the astrometric and photometric reduction use the {\sc Astro-WISE} pipeline \citep{McFarland13}; details of the resulting astrometric and photometric accuracy can be found in \citet{DeJong15}. Photometric redshifts have been derived with {\sc BPZ} \citep{Benitez00,Hildebrandt12}, after correcting the magnitudes in the optical bands for seeing differences by homogenising the photometry \citep{Kuijken15}. The photometric redshifts are reliable in the redshift range $0.005<z_{\rm B}<1.2$, with $z_{\rm B}$ being the location where the posterior redshift probability distribution has its maximum, and have a typical outlier rate of $<$5\% at $z_{\rm B}<0.8$ and a redshift scatter of 0.05 \citep[see Sect. 4.4 in][]{Kuijken15}. In the lensing analysis, we use the full photometric redshift probability distributions. \\
\indent Shear measurements are performed in the $r$-band, which has been observed under stringent seeing requirements ($<0.8''$). The $r$-band is separately reduced with the well-tested {\sc THELI} pipeline \citep{Erben05,Erben09}, following procedures very similar to the analysis of the CFHTLS data as part of the CFHTLenS collaboration \citep{Heymans12,Erben13}. The shape measurements are performed with \emph{lens}fit \citep{Miller07,Kitching08}, using the version presented in \citet{Miller13}. We apply the same calibration scheme to correct for multiplicative bias as the one employed in CFHTLenS; the accuracy of the correction is better than the current statistical uncertainties, as is shown by a number of systematics tests in \citet{Kuijken15}. Shear estimates are obtained using all source galaxies in unmasked areas with a non-zero \emph{lens}fit weight and for which the peak of the posterior redshift distribution is in the range $0.005<z_{\rm B}<1.2$. The corresponding effective source number density is 5.98 arcmin$^{-2}$ \citep[using the definition of \citet{Heymans12}, which differs from the one adopted in \citet{Chang13}, as discussed in][]{Kuijken15}. \\

\subsection{GAMA}\label{sec_data_gam}
\begin{figure}
  \includegraphics[width=1.\linewidth,angle=-90]{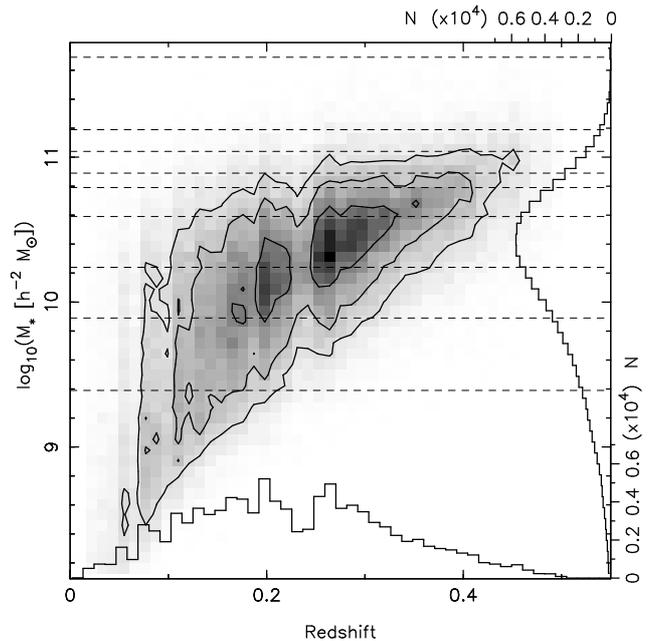}
  \caption{Spectroscopic redshift versus stellar mass of the GAMA galaxies in the KiDS overlap. The density contours are drawn at 0.5, 0.25 and 0.125 times the maximum density in this plane. The total density of GAMA galaxies as a function of redshift and stellar mass are shown by the histograms on the x- and y-axes, respectively. The dashed lines indicate the mass bins of the lenses.}
  \label{plot_GAMA_msz}
\end{figure}
The Galaxy And Mass Assembly (GAMA) survey \citep{Driver09,Driver11,Liske15} is a highly complete optical spectroscopic survey that targets galaxies with $r<19.8$ over roughly 286 deg$^2$. In this work, we make use of the G$^3$Cv7 group catalogue and version 16 of the stellar mass catalogue, which contain \mbox{$\sim$$180\,000$} objects, divided into three separate 12$\times$5 deg$^2$ patches that completely overlap with the northern stripe of KiDS. We use the subset of $\sim$$100\,000$ objects that overlaps with the 75.1 deg$^2$ from the KiDS-DR1/2. \\
\indent Stellar masses of GAMA galaxies have been estimated in \citet{Taylor11}. In short, stellar population synthesis models from \citet{Bruzual03} that assume a \citet{Chabrier03} Initial Mass Function (IMF) are fit to the $ugriz$-photometry from SDSS. NIR photometry from VIKING is used when the rest-frame wavelength is less than $11\,000$ $\AA$. To account for flux outside the AUTO aperture used for the Spectral Energy Distributions (SEDs), an aperture correction is applied using the fluxscale parameter. This parameter defines the ratio between $r$-band (AUTO) aperture flux and the total $r$-band flux determined from fitting a S\'{e}rsic profile out to 10 effective radii \citep{Kelvin12}. The stellar masses do not include the contribution from stellar remnants. The stellar mass errors are $\sim$0.1 dex and are dominated by a magnitude error floor of 0.05 mag, which is added in quadrature to all magnitude errors, thus allowing for systematic differences in the photometry between the different bands. The random errors on the stellar masses are therefore even smaller, and we ignore them in the remainder of this work. Systematic errors due to e.g. the choice of the IMF are not included in the error budget; their expected magnitude is also $\sim$0.1 dex. \\
\indent The distribution of stellar mass versus redshift of all GAMA galaxies in the KiDS footprint is shown in Fig. \ref{plot_GAMA_msz}. This figure shows that the GAMA catalogue contains galaxies with redshifts up to $z\simeq0.5$. Furthermore, bright (massive) galaxies reside at higher redshifts, as expected for a flux-limited survey. Note that the apparent lack of galaxies more massive than a few times $10^{10}h^{-2}M_\odot$ at $z<0.2$ is a consequence of plotting redshift on the horizontal axis instead of bins of equal comoving volume. The bins at low redshift contain less volume and therefore have fewer galaxies (for a constant number density). It is not a selection effect. \\
\indent We use the group properties of the G$^3$C catalogue \citep{Robotham11} to select galaxies in dense environments. Groups are found using an adaptive friends-of-friends algorithm, linking galaxies based on their projected and line-of-sight separations. The algorithm has been tested on mock catalogues, and the global properties, such as the total number of groups, are well recovered. Version 7 of the group catalogue, which we use in this work, consists of nearly 24\,000 groups with over $\sim$70\,000 group members. The catalogue contains group membership lists and various estimates for the group centre, as well as group velocity dispersions, group sizes and estimated halo masses. We limit ourselves to groups with a multiplicity $N_{\rm fof}\ge5$, because groups with fewer members are more strongly affected by interlopers, as a comparison with mock data has shown \citep{Robotham11}. We refer to these groups as `rich' groups. We assume that the brightest\footnote{ This is based on SDSS $r$-band Petrosian magnitudes with a global ($k+e$)-correction \citep[see][]{Robotham11}. Due to variations in the mass-to-light ratio, it occasionally happens that a satellite has a larger stellar mass than the central.} group galaxy is the central galaxy, whilst fainter group members are referred to as satellites. An alternative procedure to select the central galaxy is to iteratively remove group members that are furthest away from the group centre of light. As the two definitions only differ for a few percent of the groups and the lensing signals are statistically indistinguishable \citep[see Appendix A of][]{Viola15}, we do not investigate this further and adopt the brightest group galaxy as the central throughout. To verify that these `rich' groups trace dense environments, we match the G$^3$C catalogue to the environmental classification catalogue of \citet{Eardley15}, who uses a tidal tensor prescription to distinguish between four different environments: voids, sheets, filaments and knots. Using the classification that is based on the \mbox{4 $h^{-1}$Mpc} smoothing scale, we find that 76\% of the centrals of groups with $N_{\rm fof}\ge5$ reside in filaments and knots, compared to 49\% of the full GAMA catalogue, which shows the $N_{\rm fof}\ge5$ groups form a crude tracer of dense regions. \\
\indent Note that both the stellar mass catalogue and the GAMA group catalogue were derived with slightly different cosmological parameters: \citet{Taylor11} used ($\Omega_\Lambda, \Omega_{\rm M}, h$)=(0.7,0.3,0.7) and \citet{Robotham11} used ($\Omega_\Lambda, \Omega_{\rm M}, h$)=(0.75,0.25,1.0) in order to match the Millennium Simulation mocks. We accounted for the difference in $h$, but not in $\Omega_\Lambda$ and $\Omega_\rmM$, because the lensing signal at low redshift depends only weakly on these parameters and this should not impact our results. \\
\indent The current lensing catalogues in combination with these GAMA catalogues have already been analysed by \citet{Viola15}, where the main focus was GAMA group properties, and by \citet{Sifon15}, where the masses of satellites in groups were derived. Here, we aim at a broader scope, as we measure the stellar-to-halo mass relation over two orders of magnitude in halo mass. Studying the centrals and satellites in `rich' groups supplies us with the first observational limits on whether the stellar-to-halo mass relation changes in dense environments.

\subsection{Lensing signal}\label{sec_data_lens}
\begin{figure*}
  \resizebox{\hsize}{!}{\includegraphics{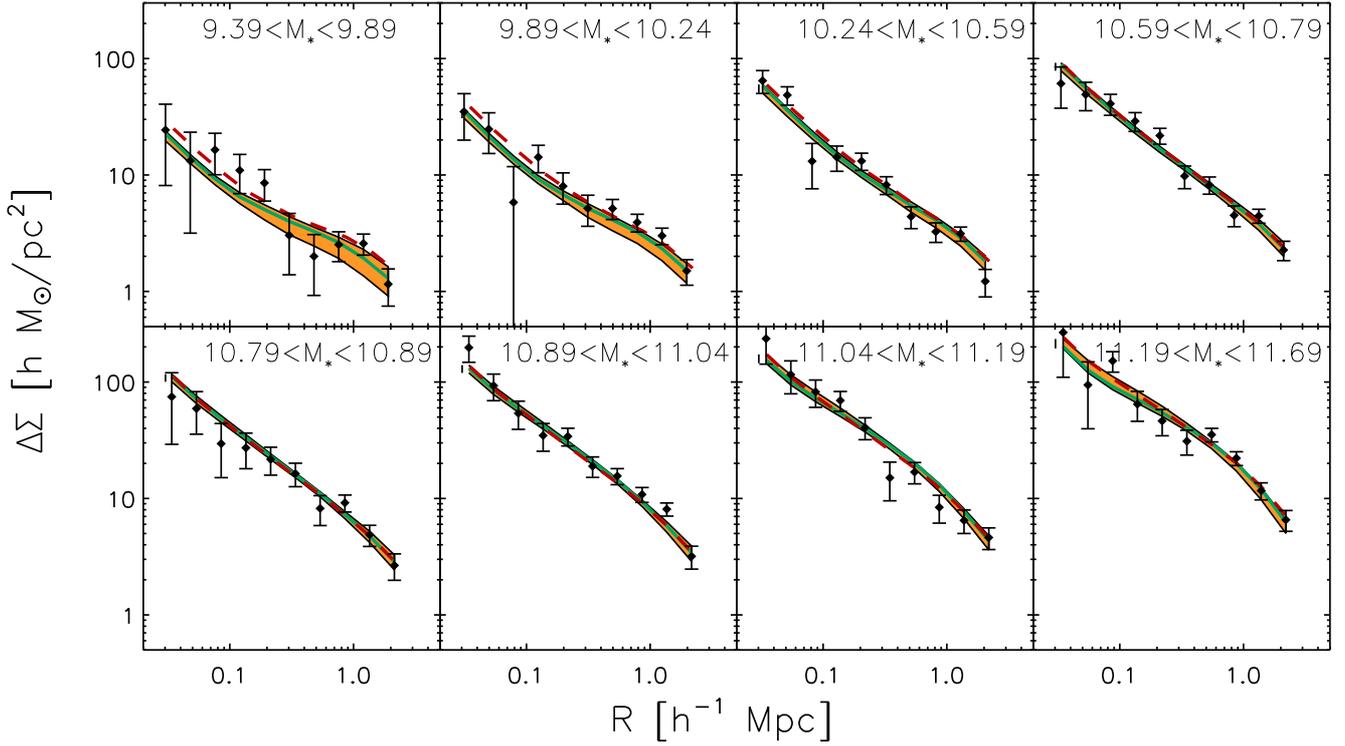}}
  \caption{Excess surface mass density profile of GAMA galaxies measured as a function of projected (comoving) separation from the lens, selected in various stellar mass bins, measured using the source galaxies from KiDS. The dashed red line indicates the best-fit halo model, obtained from fitting the lensing signal only. The solid green line is the best-fit halo model for the combined fit to the weak lensing signal and the stellar mass function, the orange area indicates the 1$\sigma$ model uncertainty regime of this fit. The stellar mass ranges that are indicated correspond to the $\log_{10}$ of the stellar masses and are in units $\log_{10}(h^{-2}M_\odot)$.}
  \label{plot_GAMA_gg_all_mstel}
\end{figure*}
Weak lensing induces a small distortion of the images of background galaxies. Since the lensing signal of individual galaxies is generally too weak to be detected due to the low number density of background galaxies in wide-field surveys, it is common practice to average the signal around many (similar) lens galaxies. In the regime where the surface mass density is sufficiently small, the lensing signal can be approximated by averaging the tangential projection of the ellipticities of background (source) galaxies, the tangential shear:
\begin{equation}
  \langle\gamma_{\rm t}\rangle(R) = \frac{\Delta\Sigma(R)}{\Sigma_{\mathrm{crit}}},
\end{equation}
with $\Delta\Sigma(R)=\bar{\Sigma}(<R)-\bar{\Sigma}(R)$ the difference between the mean projected surface mass density inside a projected radius $R$ and the surface density at $R$, and $\Sigma_{\mathrm{crit}}$ the critical surface mass density:
\begin{equation}\label{eq:sigmacit}
  \Sigma_{\mathrm{crit}}=\frac{c^2}{4\pi G}\frac{D_{\scriptscriptstyle\rm S}}{D_{\scriptscriptstyle\rm L} D_{\scriptscriptstyle\rm LS}},
\end{equation}
with $D_{\scriptscriptstyle\rm L}$ and $D_{\scriptscriptstyle\rm S}$ the angular diameter distance from the observer to the lens and source, respectively, and $D_{\scriptscriptstyle\rm LS}$ the distance between the lens and source. For each lens-source pair we compute $1/\Sigma_{\mathrm{crit}}$ by integrating over the redshift probability distribution of the source. We have not computed the error on $\Sigma_{\rm crit}$ and propagated it in the analysis. This would require knowledge on the error on the redshift probability distribution of the sources, which is not available. However, for lenses and sources that are well separated in redshift, as is the case here, the lensing efficiency $D_{\scriptscriptstyle\rm LS}/D_{\scriptscriptstyle\rm S}$ is not very sensitive to details of the source redshift distribution. Hence we expect that the error on $\Sigma_{\rm crit}$ can be safely ignored in our analysis.  \\
\indent The actual measurements of the excess surface density profiles are performed using the same methodology outlined in Sec. 3.3 of \citet{Viola15}. The covariance between the radial bins of the lensing measurements is derived analytically, as discussed in Sec. 3.4 of \citet{Viola15}. We have also computed the covariance matrix using bootstrapping techniques and found very similar results in the radial range of interest. \\
\indent We group GAMA galaxies in stellar mass bins and measure their average lensing signals. The bin ranges were chosen following two criteria. Firstly, we aimed for a roughly equal lensing signal-to-noise ratio of $\sim$15 per bin. Secondly, we adopted a maximum bin width of 0.5 dex. To determine the signal-to-noise ratio, we fitted a singular isothermal sphere (SIS) to the average lensing signal and determined the ratio of the amplitude of the SIS to its error. The adopted bin ranges are listed in Table \ref{tab_lens}, as well as the number of lenses and their average redshift; the average $\Delta\Sigma$ is shown in Fig. \ref{plot_GAMA_gg_all_mstel}. We note, however, that our conclusions do not depend on the choice of binning.
\begin{table*}
  \caption{Number of lenses and mean lens redshift of all lens samples used in this work. The stellar mass ranges that are indicated correspond to the $\log_{10}$ of the stellar masses and are in units $\log_{10}(h^{-2}M_\odot)$. The `All' sample contains all GAMA galaxies that overlap with KiDS-DR1/2, whilst `Cen' and `Sat' refers to the samples that only contain the centrals and satellites in GAMA groups with a multiplicity ${\rm N_{\rm fof}}\ge5$.}   
  \centering
  \renewcommand{\tabcolsep}{0.06cm}
  \begin{tabular}{c c c c c c c c c c c c c c c c c} 
  \hline
  & & & & & & & & & & & & & & & & \\
   & \multicolumn{2}{c}{M1} & \multicolumn{2}{c}{M2} & \multicolumn{2}{c}{M3} & \multicolumn{2}{c}{M4} & \multicolumn{2}{c}{M5} & \multicolumn{2}{c}{M6} & \multicolumn{2}{c}{M7} & \multicolumn{2}{c}{M8} \\
   & \multicolumn{2}{c}{[9.39,9.89]} & \multicolumn{2}{c}{[9.89,10.24]} & \multicolumn{2}{c}{[10.24,10.59]} & \multicolumn{2}{c}{[10.59,11.79]} & \multicolumn{2}{c}{[10.79,10.89]} & \multicolumn{2}{c}{[10.89,11.04]} & \multicolumn{2}{c}{[11.04,11.19]} & \multicolumn{2}{c}{[11.19,11.69]} \\
  \hline\hline
 & N$_{\rm{lens}}$ & $\langle z \rangle$ & N$_{\rm{lens}}$ & $\langle z \rangle$ &N$_{\rm{lens}}$ & $\langle z \rangle$ &N$_{\rm{lens}}$ & $\langle z \rangle$ &N$_{\rm{lens}}$ & $\langle z \rangle$ &N$_{\rm{lens}}$ & $\langle z \rangle$ &N$_{\rm{lens}}$ & $\langle z \rangle$ &N$_{\rm{lens}}$ & $\langle z \rangle$ \\
   & & & & & & & & & & & & & & & & \\
 All & 15819 & 0.17 & 19175 & 0.21 & 24459 & 0.25 & 11475 & 0.29 & 3976 & 0.31 & 3885 & 0.32 & 1894 & 0.34 & 1143 & 0.35 \\
 Cen (${N_{\rm fof}}\ge5$) & 15 & 0.08 & 55 & 0.12 & 185 & 0.16 & 242 & 0.18 & 185 & 0.19 & 276 & 0.21 & 241 & 0.23 & 209 & 0.26 \\
 Sat (${N_{\rm fof}}\ge5$)  & 1755 & 0.14 & 2392 & 0.18 & 3002 & 0.22 & 1267 & 0.26 & 388 & 0.27 & 343 & 0.27 & 138 & 0.29 & 65 & 0.32 \\
   & & & & & & & & & & & & & & & & \\
  \hline \\
  \end{tabular}
  \label{tab_lens}
\end{table*} 

\subsection{The halo model}\label{sec_data_hm}
The halo model \citep{Seljak00,CoorayS02} has become a standard method to interpret weak lensing data. The implementation we employ here is similar to the one described in \citet{Vandenbosch13} and has been successfully applied to weak lensing measurements in \citet{Cacciato14,VanUitert15}, and to weak lensing and galaxy clustering data in \citet{Cacciato13}. We provide a description of the model here, as we have made a number of modifications. \\
\indent In the halo model, all galaxies are assumed to reside in spherical dark matter haloes. Using a prescription for the way galaxies occupy dark matter haloes, as well as for the matter density profile, abundance and clustering of haloes, one can predict the surface mass density (and thus the lensing signal) correlated with galaxies in a statistical manner:
\begin{equation}
\Sigma(R) = \bar{\rho}_{\rm m} \int_{0}^{\omega_{\scriptscriptstyle\rm S}} \xi_{\rm gm}(r) {\rm d} \omega,
\label{Sigma_approx}
\end{equation}
with  $\xi_{\rm gm}(r)$ the galaxy-matter cross-correlation, $\omega$ the comoving distance from the observer and $\omega_{\scriptscriptstyle\rm S}$ the comoving distance to the source. For small separations, $R \approx \omega_{\scriptscriptstyle\rm L}\, \theta$, with $\omega_{\scriptscriptstyle\rm L}$ the comoving distance to the lens and $\theta$ the angular separation from the lens. The three-dimensional comoving distance $r$ is related to $\omega$ via $r^2 = (\omega_{\scriptscriptstyle\rm L}\cdot \theta)^2  + (\omega - \omega_{\scriptscriptstyle\rm L})^2$. The integral is computed along the line of sight. \\
\indent As the computation of $\xi_{\rm gm}(r)$ generally requires convolutions in real space, it is convenient to express the relevant quantities in Fourier-space where these operations become multiplications. $\xi_{\rm gm}(r)$ is related to the galaxy-matter power spectrum, $P_{\rm gm}(k,z)$, via
\begin{eqnarray}
\xi_{\rm gm}(r,z) = \frac{1}{2 \pi^2} \int_0^{\infty} P_{\rm gm}(k,z)
\frac{\sin kr}{kr} k^2 {\rm d} k,
\label{xiFTfromPK}
\end{eqnarray}
with $k$ the wavenumber. On small physical scales, the main contribution to $P_{\rm gm}(k,z)$ comes from the halo in which a galaxy resides (the one-halo term), whilst on large physical scales, the main contribution comes from neighbouring haloes (the two-halo term). Additionally, the halo model distinguishes between two galaxy types, i.e. centrals and satellites. Centrals reside in the centre of a main halo, whilst satellites reside in subhaloes that are embedded in larger haloes. Their power spectra are different and computed separately. Hence one has
\begin{equation}
P_{\rm gm}(k) = P^{\rm 1h}_{\rm cm}(k) + P^{\rm 1h}_{\rm sm}(k) +P^{\rm 2h}_{\rm cm}(k) + P^{\rm 2h}_{\rm sm}(k),
\label{eq_powerspec}
\end{equation}
with $P^{\rm 1h}_{\rm cm}(k)$ ($P^{\rm 1h}_{\rm sm}(k)$) the one-halo contributions from centrals (satellites), and $P^{\rm 2h}_{\rm cm}(k)$ ($P^{\rm 2h}_{\rm sm}(k)$) the corresponding two-halo terms. We follow the notation of \citet{Vandenbosch13} and write this compactly as:
\begin{equation}
P^{\rm 1h}_{\rm xy}(k,z) = \int \calH_\rmx(k,M_\rmh,z)  \calH_\rmy(k,M_\rmh,z) n_{\rm h}(M_\rmh,z) {\rm d} M_\rmh,
\label{P1h}
\end{equation}
\begin{eqnarray}
\lefteqn{P^{\rm 2h}_{\rmx\rmy}(k,z) = \int {\rm d} M_1 \calH_{\rm x}(k,M_1,z)  n_{\rm h}(M_1,z) } \nonumber \\
& & \int {\rm d} M_2  \calH_\rmy(k,M_2,z)  n_{\rm h}(M_2,z) Q(k|M_1,M_2,z),
\label{eq_2h}
\end{eqnarray}
where x and y are either c (for central), s (for satellite), or m (for matter), $n_{\rm h}(M_\rmh,z)$ is the halo mass function of \cite{Tinker10}, and \mbox{$Q(k|M_1,M_2,z)= b_\rmh(M_1,z)b_\rmh(M_2,z)P^{\rm lin}_{\rm m}(k,z)$} describes the power spectrum of haloes of mass $M_1$ and $M_2$,  which contains the large-scale halo bias $b_\rmh(M_\rmh)$ from \cite{Tinker10}. $P^{\rm lin}_{\rm m}(k,z)$ is the linear matter power spectrum. We employ the transfer function of \citet{Eisenstein98}, which properly accounts for the acoustic oscillations. Furthermore, we use
\begin{equation}
\calH_{\rm m}(k,M_\rmh,z) = \frac{M_\rmh}{\bar{\rho}_{\rm m}} \widetilde{u}_{\rm h}(k|M_\rmh,z),
\label{calHm}
\end{equation}
with $M_\rmh$ the halo mass, and $\widetilde{u}_{\rm h}(k|M_\rmh,z)$ the Fourier transform of the normalised density distribution of the halo. We assume that the density distribution follows a Navarro-Frenk-White \citep[NFW;][]{Navarro96} profile, with a mass-concentration relation from \citet{Duffy08}:
\begin{equation}
c_{\rm dm} = f_{\rm conc} \times 10.14\left(\frac{M_\rmh}{M_{\rm pivot}}\right)^{-0.081} (1+z)^{-1.01}\, ,
\label{eq_mc}
\end{equation}
where $f_{\rm conc}$ is the normalization, which is a free parameter in the fit, and $M_{\rm pivot}=2\times10^{12}h^{-1}M_\odot$. Note that the choice for this particular parametrisation is not very important, as essentially all mass-concentration relations from the literature predict a weak dependence on halo mass. Furthermore, the scaling with redshift $c_{\rm dm} \propto (1+z)^{-1}$ is motivated by analytical treatments of halo formation \citep[see e.g.][]{Bullock01}. It is worth mentioning that more complex redshift dependences are expected \citep[see e.g.][]{Munoz11} but those deviations are only relevant at redshifts larger than one, well beyond the highest lens redshift in this study.\\
\indent For centrals and satellites, we have
\begin{equation}
\calH_{\rm x}(k,M_\rmh,z) = \frac{\langle N_{\rm x}|M_\rmh \rangle}{\bar{n}_{\rmx}(z)} {\widetilde u}_{\rm x}(k|M_\rmh).
\label{calHc}
\end{equation}
We set ${\widetilde u}_{\rm c}(k|M_\rmh) = 1$, i.e., we assume that all central galaxies are located at the centre of the halo. We adopt this choice in order to limit the number of free parameters in the model; additionally, lensing alone does not provide tight constraints on the miscentring distribution. This modelling choice can lead to a biased normalisation of the mass-concentration relation \citep[see e.g.][]{VanUitert15,Viola15} but it does not bias the halo masses \citep{VanUitert15} or the stellar-to-halo mass relation. Furthermore, we assume \mbox{$\widetilde{u}_{\rm s}(k|M_\rmh,z) = \widetilde{u}_{\rm h}(k|M_\rmh,z)$}, hence the distribution of satellites follows the dark matter. This is a reasonable assumption, given the large discrepancies in the reported trends in the literature, which range from  satellites being either more or less concentrated than the dark matter \citep[see e.g.][and the discussion therein]{Wang14}. \\
\indent We specify the halo occupation statistics using the Conditional Stellar Mass Function (CSMF), $\Phi(M_*|M_\rmh)\rmd M_*$, which describes the average number of galaxies with stellar masses in the range $M_* \pm \rmd  M_*/2$ that reside in  a halo of mass $M_\rmh$. The occupation numbers required for the computation of the galaxy-matter power spectra follow from
\begin{equation}\label{eq:Ns}
\langle N_{\rm x}| M_\rmh \rangle(M_{*,1},M_{*,2}) =
\int_{M_{*,1}}^{M{*,_2}} \Phi_{\rm x}(M_*|M_\rmh) \rmd M_*\,,
\end{equation}
where   `x'  refers to either   `c'   (centrals) or `s'  (satellites), and $M_{*,1}$ and $M_{*,2}$ indicate the extremes of a stellar mass bin. The average number density of these galaxies is given by:
\begin{equation}
\bar{n}_{\rmx}(z)=\int \langle N_\rmx | M_\rmh \rangle(M_{*,1},M_{*,2}) n_{\rm h}(M_\rmh,z) {\rm d} M_\rmh,
\end{equation}
and the satellite fraction follows from
\begin{equation}
f_\rms(M_{*,1},M_{*,2}) = \frac{\int \langle N_{\rm s}| M_\rmh \rangle(M_{*,1},M_{*,2}) \, n_{\rm h}(M_\rmh) \, {\rm d}M_\rmh \,}{\bar{n}_{\rmc}(z)+\bar{n}_{\rms}(z)} \, .
\label{eq_fsat}
\end{equation}
The stellar mass function is given by:
\begin{equation}
\varphi(M_{*,1},M_{*,2}) = \int [\langle N_{\rm c}| M_\rmh \rangle + \langle N_{\rm s}| M_\rmh \rangle]  \, n_{\rm h}(M_\rmh) \, \rmd M_\rmh\, ,
\label{eq_smf}
\end{equation}
where $\langle N_{\rm c}| M_\rmh \rangle$ and $\langle N_{\rm s}| M_\rmh \rangle$ are computed with Eq. (\ref{eq:Ns}) using the bin limits of the stellar mass function. \\
\indent We separate the CSMF into the contributions of central and satellite galaxies, \mbox{$\Phi(M_*|M_\rmh) = \Phi_\rmc(M_*|M_\rmh) + \Phi_\rms(M_*|M_\rmh)$}. The contribution from the central galaxies is modelled as a log-normal distribution:
\begin{equation}\label{eq:phi_c}
\Phi_\rmc(M_*|M_\rmh) =  \frac{\exp \left[- { {(\log_{10} M_*  -\log_{10} M^\rmc_*(M_\rmh) )^2 } \over 2\sigma_\rmc^2} \right]}{\sqrt{2\pi} \, {\rm ln}(10) \, \sigma_\rmc \, M_*} \, ,
\end{equation}
where $\sigma_\rmc$ is the scatter in $\log M_*$ at a fixed halo mass. For simplicity, we assume that it does not vary with halo mass, as supported by the kinematics of satellite galaxies in the SDSS \citep{More09,More11}, by combining galaxy clustering, galaxy-galaxy lensing and galaxy abundances \citep{Cacciato09,Leauthaud12} and by SDSS galaxy group catalogues \citep{Yang08}. $M^\rmc_*$ represents the mean stellar mass of central galaxies in a halo of mass $M_\rmh$, parametrised by a double power law:
\begin{equation}
M^\rmc_*(M_\rmh) = M_{*,0} {(M_\rmh/M_{\rmh,1})^{\beta_1} \over 
\left[1 + (M_\rmh/M_{\rmh,1}) \right]^{\beta_1-\beta_2}}\,,
\label{eq_msmh}
\end{equation}
with $M_{\rmh,1}$ a  characteristic mass scale, $M_{*,0}$ a normalization and $\beta_1$ ($\beta_2$) the power law slope at the low-(high-) mass end. This is the stellar-to-halo mass relation of central galaxies we are after.\\
\indent For the CSMF of the satellite galaxies, we adopt a modified Schechter function:
\begin{equation}
\Phi_\rms(M_*|M_\rmh) = {\phi_\rms\over M^\rms_*} 
\left({M_*\over M^\rms_*}\right)^{\alpha_\rms}
{\rm exp} \left[- \left ({M_*\over M^\rms_*} \right )^2 \right],
\label{eq_phis}
\end{equation}
which decreases faster than a Schechter function at the high-stellar mass end. Galaxy group catalogues show that the satellite contribution to the total CSMF falls off around the mean stellar mass of the central galaxy for a given halo mass \citep[e.g.][]{Yang08}. Thus one expects the characteristic mass of the modified Schechter function, $ M^\rms_*$, to follow $M^\rmc_*$. Inspired by \citet{Yang08}, we assume that $M^\rms_*(M_\rmh)  = 0.56 M^\rmc_*(M_\rmh)$. For the normalization of $\Phi_\rms(M_*|M_\rmh)$ we adopt
\begin{equation}\label{eq:phi}
\log_{10}[\phi_\rms(M_\rmh)] = b_0 + b_1 \times \log_{10} M_{13}\,,
\end{equation}
with $M_{13}=M_\rmh/(10^{13}  h^{-1}\Msun)$. $b_0$, $b_1$, and  $\alpha_\rms$ are free parameters. We test the sensitivity of our results to the location of $M^\rms_*$, and to the addition of a quadratic term in Eq. (\ref{eq:phi}), in Appendix \ref{app_sens}. We find that our results are not significantly affected. \\
\indent We assign a mass to the subhaloes in which the satellites reside using the same relation that we use for the centrals (Eq. \ref{eq_msmh}). For every stellar mass bin, we compute the average mass of the main haloes in which the satellite resides, and multiply this with a constant factor, $f_{\rm sub}$, a free parameter whose range is limited to values between 0 and 1. In this way, we can crudely account for the stripping of the dark matter haloes of satellites. Given our limited knowledge of the distribution of dark matter in satellite galaxies, we assume that it is described by an NFW profile, which provide a decent description of the mass distribution of subhaloes in the Millennium simulation \citep{Pastor11}. We use the same mass-concentration relation as for the centrals. The statistical power of our measurements is not sufficient to additionally fit for a truncation radius \citep[see][]{Sifon15}. \\
\indent The above prescription provides us with the lensing signal from centrals and satellites, with separate contributions from their one-halo and two-halo terms. At small projected separations, the contribution of the baryonic component of the lenses themselves becomes relevant. We model this using a simple point mass approximation:
\begin{equation}
\Delta \Sigma^{\rm 1h}_{\rm gal}(R) \equiv \frac{\langle M_* \rangle }{\pi \, R^2}\, ,
\end{equation}
with $\langle M_* \rangle$ the average stellar mass of the lens sample.  \\
\indent To summarise, the halo model employed in this paper has the following free parameters: $(M_{\rmh,1}, M_{*,0}, \beta_1, \beta_2, \sigma_\rmc)$ and $(\alpha_{\rm s},b_0,b_1)$ to describe the halo occupation statistics of centrals and satellites. $f_{\rm sub}$ controls the subhalo masses of satellites, and $f_{\rm conc}$ quantifies the normalization of the $c(M)$ relation. We use non-informative flat or Gaussian priors, as listed in Table \ref{tab_prior}, except for $\beta_1$, because our measurements do not extend far below the location of the kink in the stellar-to-halo mass relation. As a consequence, we are not able to provide tight constraints on the slope at the low-mass end. All priors were chosen to generously encapsulate previous literature results and they do not affect our results. In particular, in Appendix \ref{app_sens} we demonstrate that our results are insensitive to the choice of prior on $\beta_1$. Note that for some samples, we had to adopt somewhat different priors; we comment on this where applicable. \\
\indent The parameter space is sampled with an affine invariant ensemble Markov Chain Monte Carlo (MCMC) sampler \citep{Goodman10}. Specifically, we use the publicly available code {\sc Emcee} \citep{Foreman13}. We run {\sc Emcee} with four separate chains with 150 walkers and 4\,500 steps per walker. The first 1000 steps (which amounts to 600\,000 evaluations) are discarded as the burn-in phase. Using the resulting \mbox{2\,100\,000} model evaluations, we estimate the parameter uncertainties; the fit parameters that we quote in the following correspond to the median of the marginalized posterior distributions, the errors correspond to the 68\% confidence intervals around the median. We assess the convergence of the chains with the Gelman-Rubin test \citep{Gelman92} and ensure that $R\leq1.015$, with $R$ the ratio between the variance of a parameter in the single chains and the variance of that parameter in all chains combined. In addition, we compute the auto-correlation time \citep[see e.g.][]{Akaret13} for our main results and find that it is shorter than the length of the chains that is needed to reach 1\% precision on the mean of each fit parameter. \\
\indent For some lens selections, we also run the halo model in an `informed' setting. When we use the GAMA group catalogue to select and analyse only centrals or satellites, we only need the part of the halo model that describes their respective signals. Hence, when we only select centrals, we set the CSMF of satellites to zero. When we select satellites only, we model both the CSMF of the satellites and of the centrals of the haloes that host the satellites. We need the latter to model the miscentred one-halo term and the subhalo masses of the satellites. \\
\begin{figure*}
  \begin{minipage}{0.48\linewidth}
  \includegraphics[width=1.\linewidth]{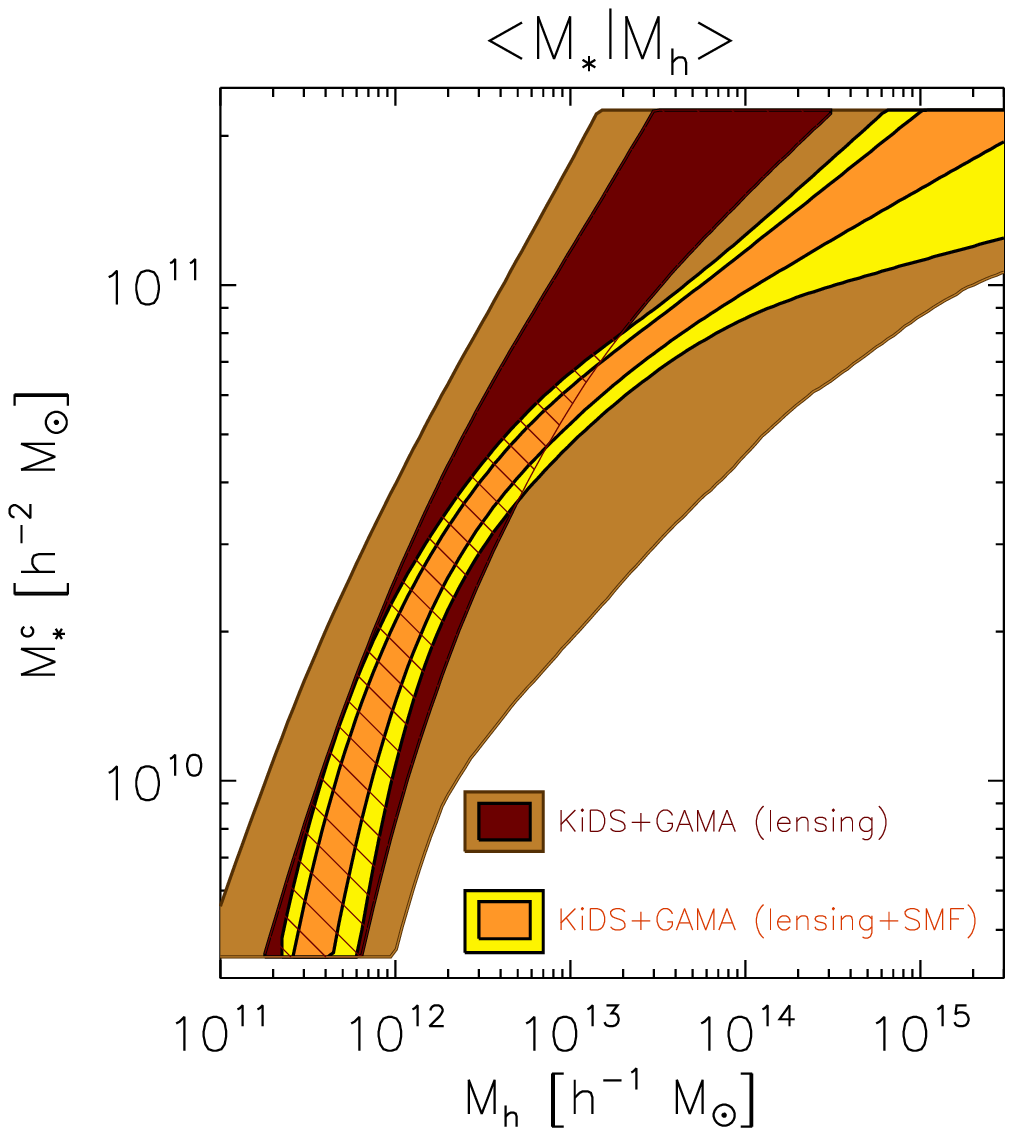}
  \end{minipage}
  \begin{minipage}{0.48\linewidth}
  \includegraphics[width=1.\linewidth]{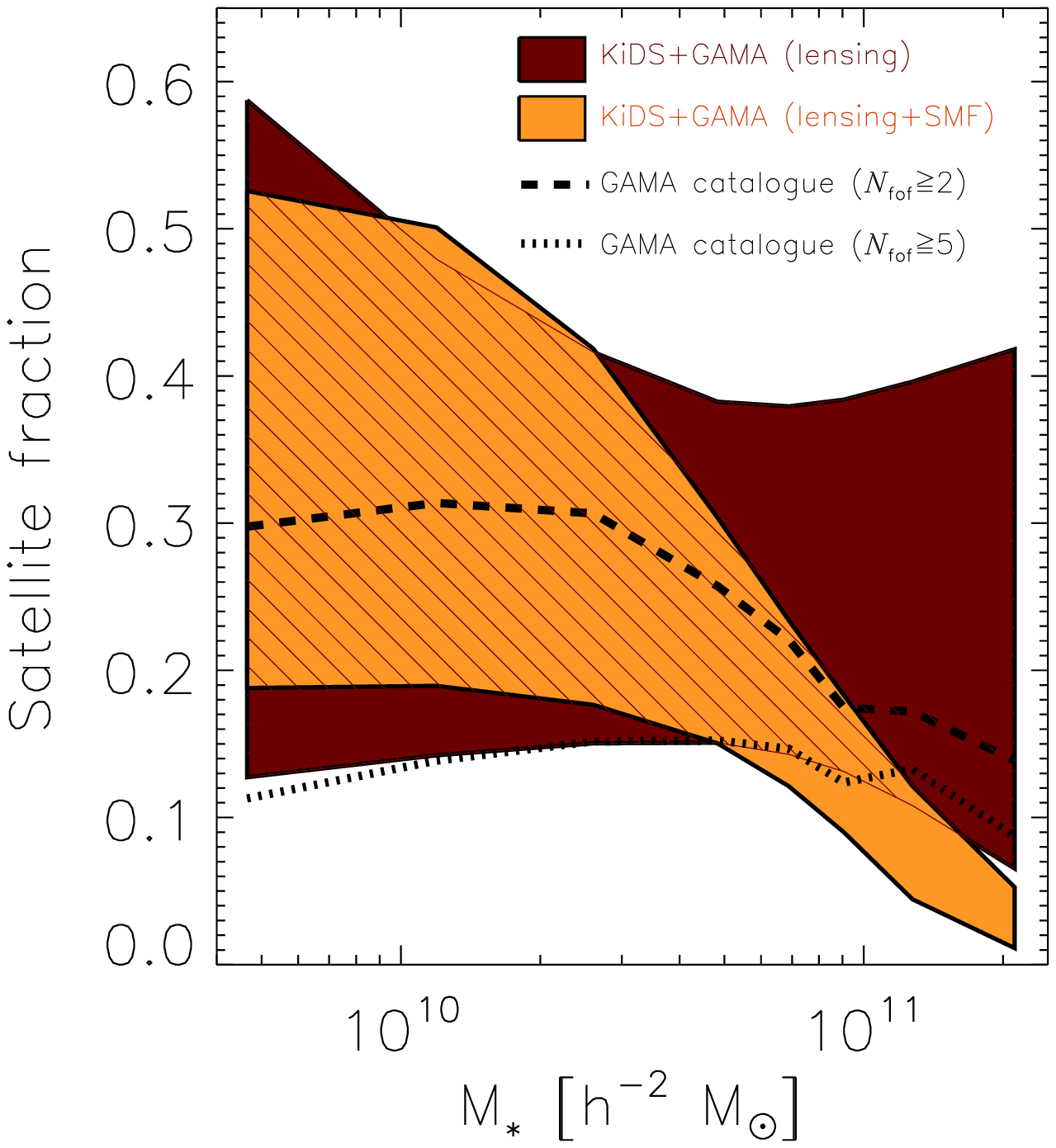}
  \end{minipage}
  \caption{{\it (left:)} Stellar-to-halo mass relation of central galaxies from KiDS+GAMA, determined from fitting the lensing signal only (dark/light brown indicating 1-/2-$\sigma$ model uncertainty regime) or by combining the lensing signal with the stellar mass function (orange/yellow indicating 1-/2-$\sigma$ model uncertainty regime). The contours are cut at the mean stellar mass of the first and last stellar mass bin used in the lensing analysis, to ensure we only show the regime where the data constrains it. {\it (right:)} The fraction of galaxies that are satellites as a function of stellar mass for all GAMA galaxies. The coloured contours show the 68\% confidence interval for the fits to the lensing signal only and to the combined fits, as indicated in the panel. The upper thick black dashed line shows a crude estimate of the satellite fraction based on the GAMA group catalogues (as detailed in Sect. \ref{sec_satfrac}), the lower thick dotted line shows a lower limit. Hatched areas show the overlap between the 1-$\sigma$ lensing-only results and the combined analysis.}
  \label{plot_fsat}
\end{figure*}
\begin{table}
  \renewcommand{\tabcolsep}{0.15cm}
  \caption{Priors adopted in halo model fit}   
  \begin{tabular}{c c c c c} 
  \hline
 Parameter & type & range & prior mean & prior sigma \\
  \hline\hline
 $\log_{10}(M_{\rmh,1})$ & flat & $[9,14]$ & - & - \\
 $\log_{10}(M_{*,0})$ & flat & $[7,13]$ & - & - \\
 $\beta_1$ & Gaussian & - & 5.0 & 3.0 \\
 $\log_{10}\beta_2$ & flat & $[-3,\infty]$ & - & - \\
 $\sigma_\rmc$ & flat & $[0.05,0.5]$ & - & - \\
 $\alpha_{\rm s}$ & Gaussian & - & -1.1 & 0.9 \\
 $b_0$ & Gaussian & - & 0.0 & 1.5 \\
 $b_1$ & Gaussian & - & 1.5 & 1.5 \\
 $f_{\rm sub}$ & flat & $[0,1]$ & - & - \\
 $f_{\rm conc}$ & flat & $[0.2,2]$ & - & - \\
 $c_0$ & flat & $[-5,5]$ & - & - \\
 $c_1$ & flat & $[9,16]$ & - & - \\
  \hline
  \end{tabular}
  \label{tab_prior}
\end{table} 

\section{Stellar-to-halo mass relation}\label{sec_msmh}

\begin{figure}
  \resizebox{\hsize}{!}{\includegraphics{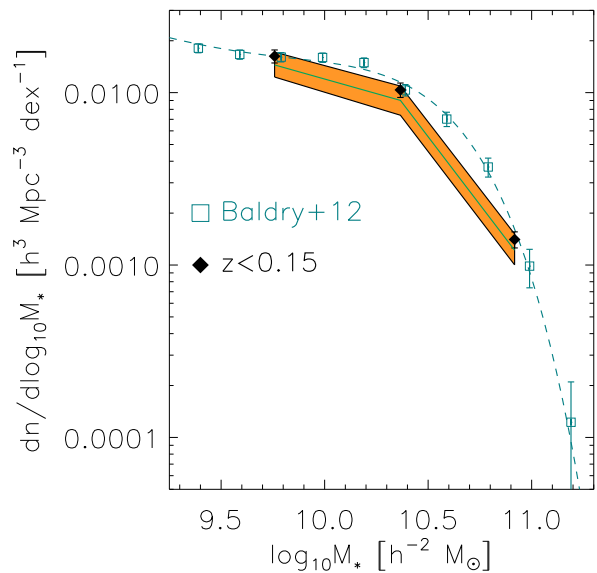}}
  \caption{Stellar mass function determined using all GAMA galaxies at $z<0.15$. Orange regions indicate the 68\% confidence interval from the halo model fit to the lensing signal and the stellar mass function, linearly interpolated between the stellar mass bins. The solid green line indicates the best fit model. The stellar mass function from \citet{Baldry12}, determined using GAMA galaxies at $z<0.06$, is also shown.}
  \label{plot_SMF}
\end{figure}

\indent We start with an analysis of the lensing measurements to examine the stellar-to-halo mass relation of central galaxies, as was done in several previous studies \citep[e.g.][]{Mandelbaum06,VanUitert11,Velander14}. We fit the lensing signals of the eight lens samples simultaneously with the halo model. The best-fitting models from the lensing-only analysis can be compared to the data in Fig. \ref{plot_GAMA_gg_all_mstel}. The resulting reduced chi-squared, $\chi^2_{\rm red}$, has a value of 1.0 (with 70 degrees of freedom), so the models provide a satisfactory fit. In the left-hand panel of Fig. \ref{plot_fsat} we show the constraints on the stellar-to-halo mass relation. A broad range of relations describe the lensing signals equally well. Furthermore, the right-hand panel of Fig. \ref{plot_fsat} shows that the uncertainties on the fraction of galaxies that are satellites is also large. \\
\indent \citet{VanUitert11} pointed out that the uncertainties on the satellite fraction obtained from lensing only are large at the high stellar mass end, and, even worse, that a wrongly inferred satellite fraction can bias the halo mass as they are anti-correlated. The reason for this degeneracy is that lowering the halo mass reduces the model excess surface mass density profile, which can be partly compensated by increasing the satellite fraction, as satellites reside on average in more massive haloes than centrals of the same stellar mass, thereby boosting the model excess surface mass density profile at a few hundred kpc. This problem was partly mitigated in \citet{VanUitert11} and \citet{Velander14} by imposing priors on the satellite fractions, which is not ideal, as the results are sensitive to the priors used. As we employ a more flexible halo model here, this problem is exacerbated and a different solution is required. \\
\indent In order to tighten the constraints on the satellite fraction and the stellar-to-halo mass relation, we either need to impose priors in the halo model, or include additional, complementary data sets. Since it is not obvious what priors to use, particularly since we aim to study how the stellar-to-halo mass relation depends on environment, we opt for the second approach. The most straightforward complementary data set is the stellar mass function, which constrains the central and satellite CSMFs through Eq. (\ref{eq_smf}). As a result, the number of satellites cannot be scaled arbitrarily up or down anymore, which helps to break this degeneracy. \\
\indent Since GAMA is a highly complete spectroscopic survey, we can measure the stellar mass function by simply counting galaxies, as long as we restrict ourselves to stellar mass and redshift ranges where the sample is volume limited. Hence we measure the stellar mass function in three equally log-spaced bins between \mbox{$9.39<\log_{10}(M_*/h^{-2}M_\odot)<11.69$} and include all galaxies from the G$^3$Cv7 group catalogue with $z<0.15$. The choice of the number of bins is mainly driven by the low number of independent bootstrap realisations we can use to estimate the errors (discussed in Appendix \ref{app_smfcovar}). We do not expect to lose much constraining power from the stellar mass function by measuring it in three bins only. Note that the mean lens redshift is somewhat higher than the redshift at which we determine the stellar mass function, but the evolution of the stellar mass function is very small over the redshift range considered in this work \citep[see, e.g.,][]{Ilbert13} and hence can be safely ignored. The stellar mass function is shown in Fig. \ref{plot_SMF}, together with the results from \citet{Baldry12}, who measured the stellar mass function for GAMA using galaxies at $z<0.06$. The measurements agree well. \\
\indent We determine the error and the covariance matrix via bootstrapping, as detailed in Appendix \ref{app_smfcovar}. We show there that 1) the bootstrap samples should contain a sufficiently large physical volume. If the sample volume is too small, the errors will be underestimated; 2) the major contribution to the error budget comes from cosmic variance. The contribution from Poisson noise is typically of order 10-20\%; 3) the stellar mass function measurements are highly correlated. \citet{Smith12} showed that this has a major impact on the confidence contours of model parameters fitted to the stellar mass function. Including the covariance is therefore essential, not only for studies that characterise the stellar mass/luminosity function (for example as a function of galaxy type), but also when it is used to constrain halo model fits. \\
\indent To determine the cross-covariance between the shear measurements and the stellar mass function, we measured the shear of all GAMA galaxies with \mbox{$\log_{10}(M_*/h^{-2}M_\odot)>9.39$} and \mbox{$z<0.15$} in each KiDS pointing, and used the same GAMA galaxies to determine the stellar mass function. These measurements were used as input to our bootstrap analysis. The covariance matrix of the combined shear and stellar mass function measurements revealed that the cross-covariance between the two probes is negligible and can be safely ignored. The covariance between the lensing measurements and the stellar mass function for smaller subsamples of GAMA galaxies is expected to be even smaller because of larger measurement noise. Therefore, we do not restrict ourselves to the overlapping area with KiDS, but use the entire 180 deg$^2$ of GAMA area to determine the stellar mass function to improve our statistics. \\
\indent We fit the lensing signal of all bins and the stellar mass function simultaneously with the halo model. The best-fit models are shown in Fig. \ref{plot_GAMA_gg_all_mstel} and  \ref{plot_SMF}, together with the 1$\sigma$ model uncertainties. The reduced $\chi^2$ of the best-fit model is $80/(83 - 10)=1.1$ (eight mass bins times ten angular bins for the lensing signal, plus three mass bins for the stellar mass function), so the halo model provides an appropriate fit. The lensing signal of the best-fit model is virtually  indistinguishable from the best-fit model of the lensing-only fit. The stellar-to-halo mass relation, however, is better constrained, as is shown in Fig. \ref{plot_fsat}. The relation is flatter towards the high mass end, as a result of a better constrained satellite fraction that decreases with stellar mass (discussed in Sect. \ref{sec_satfrac}). \\
\begin{figure*}
  \resizebox{\hsize}{!}{\includegraphics{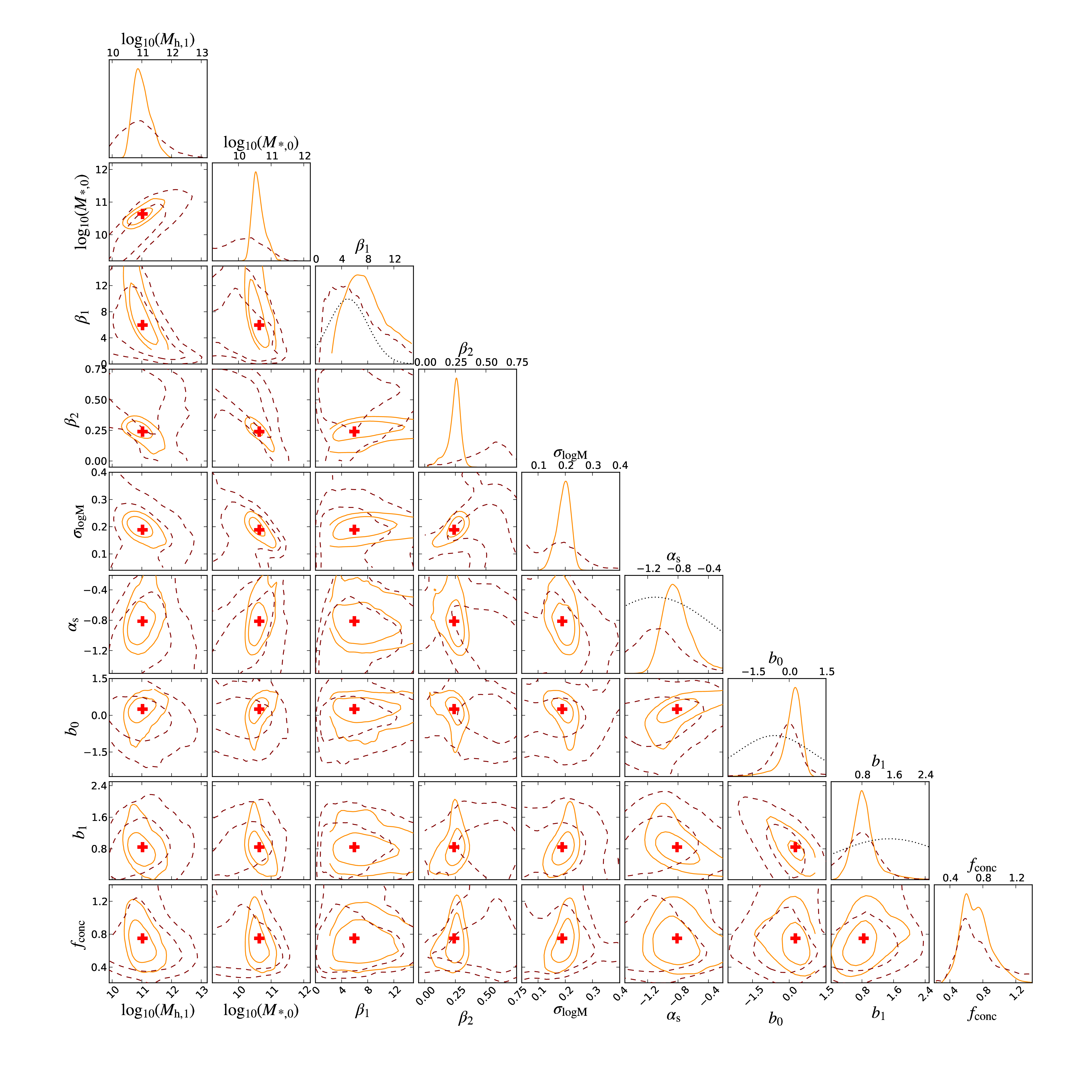}}
  \caption{Posteriors of pairs of parameters, marginalised over all other parameters. Dimensions are the same as in Table \ref{tab_res}. Solid orange contours indicate the 1-/2-$\sigma$ confidence intervals of the lensing+SMF fit, whilst the brown dashed contours indicate  the 1-/2-$\sigma$ confidence intervals of the lensing-only fit. Red crosses indicate the best-fit solution of the combined fit. The panels on the diagonal show the marginalised posterior of the individual fit parameters, together with the priors (blue dotted lines). Including the SMF in the fit mainly helps to constrain the stellar-to-halo mass relation parameters (Eq. \ref{eq_msmh}) and $\alpha_s$. The degeneracies between the stellar-to-halo mass relation parameters and those that describe the satellite CSMF (Eq. \ref{eq_phis}, \ref{eq:phi}), follow from the functional form we adopted.}
  \label{plot_post}
\end{figure*}
\begin{table*}
  \caption{Fit parameters of the halo model. Parameters (1) to (5) determine the CSMF of centrals, (6) to (8) the CSMF of satellites, (9) the subhalo masses of satellites, (10) the normalization of the mass-concentration relation, and (10) and (11) account for the selection incompleteness of centrals, as defined in Eq. (\ref{eq_incompl}). Dimensions of parameter (1) and (2) are $[\log_{10} (h^{-1}M_\odot)]$ and $[\log_{10} (h^{-2}M_\odot)]$, respectively. `Cen' and `Sat' refer to the fits to the samples that consist only of centrals and satellites in `rich' groups ($N_{\rm fof}\geq5$), respectively. A `$\star$' indicates the parameters that were fixed in the fit to the best-fit values of the halo model run on centrals only, whilst a `-' indicates that the parameter was not used in the halo model. For the satellites, we show both the results where we fixed the stellar-to-halo mass relation of the central galaxies (third row) and where we fit for it (fourth row).}   
  \centering
  \renewcommand{\tabcolsep}{0.04cm}
  \begin{tabular}{c c c c c c c c c c c c c} 
  \hline
   & & & & & & & & & & & & \\
   & $\log_{10}(M_{\rmh,1})$ & $\log_{10}(M_{*,0})$ & $\beta_1$ & $\beta_2$ & $\sigma_{\rm c}$ & $b_0$ & $b_1$ & $\alpha_s$ & $f_{\rm sub}$ & $f_{\rm conc}$ & $c_0$ & $c_1$ \\
   & & & & & & & & & & & & \\
   & (1) & (2) & (3) & (4) & (5) & (6) & (7) & (8) & (9) & (10) & (11) & (12)\\
  \hline\hline
   & & & & & & & & & & & & \\
All & $10.97^{+0.34}_{-0.25}$ &  $10.58^{+0.22}_{-0.15}$ & $7.5^{+3.8}_{-2.7}$& $0.25^{+0.04}_{-0.06}$ & $0.20^{+0.02}_{-0.03}$ & $0.18^{+0.28}_{-0.39}$ & $0.83^{+0.27}_{-0.23}$ & $-0.83^{+0.22}_{-0.16}$ & $0.59^{+0.31}_{-0.40}$ & $0.70^{+0.19}_{-0.15}$ & - & - \\
Cen & $12.06^{+0.72}_{-0.80}$ & $11.16^{+0.40}_{-0.62}$ & $5.4^{+5.3}_{-3.4}$ & $0.15^{+0.31}_{-0.14}$ & $0.14^{+0.08}_{-0.05}$  & - & - & - & - & $0.77^{+0.27}_{-0.18}$ & $2.05^{+1.88}_{-0.82}$ & $13.00^{+0.28}_{-0.13}$\\
Sat & $\star$ & $\star$ & $\star$ & $\star$ & $\star$ & $0.12^{+0.19}_{-0.26}$ & $0.71^{+0.12}_{-0.13}$ & $-1.03^{+0.07}_{-0.08}$ & $0.25^{+0.09}_{-0.08}$ & $0.94^{+0.18}_{-0.16}$ & $\star$ & $\star$ \\
Sat & $11.70^{+0.70}_{-0.84}$ & $11.22^{+0.12}_{-0.22}$ & $4.5^{+4.6}_{-2.9}$ & $0.05^{+0.07}_{-0.04}$ & $0.12^{+0.12}_{-0.05}$ & $-0.14^{+0.63}_{-0.28}$ & $1.03^{+0.14}_{-0.33}$ & $-1.00^{+0.10}_{-0.12}$ & $0.40^{+0.43}_{-0.21}$ & $1.05^{+0.25}_{-0.18}$ & $1.03^{+2.89}_{-1.01}$ & $12.09^{+2.71}_{-1.86}$ \\
  \hline \\
  \end{tabular}
  \label{tab_res}
\end{table*} 
\begin{figure*}
  \resizebox{\hsize}{!}{\includegraphics{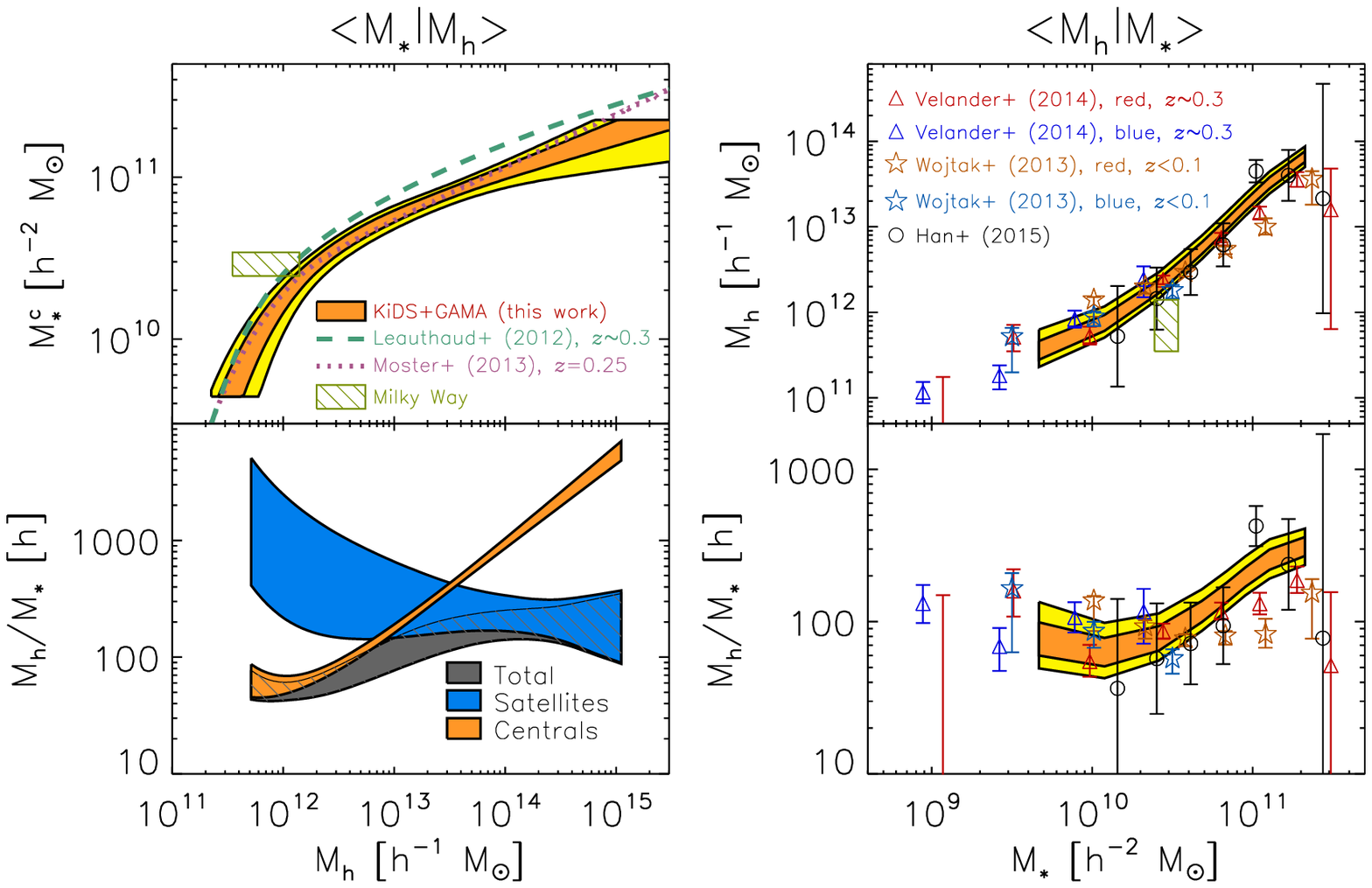}}
  \caption{Stellar-to-halo mass relation of central galaxies for KiDS+GAMA. Orange (yellow) regions indicate the 68\% (95\%) confidence intervals for the centrals, blue regions the 68\% confidence intervals for the satellites, and grey (solid and hatched) regions are the 68\% confidence intervals for the total sample. Our results can be compared to constraints from \citet{Leauthaud12}, \citet{Moster13}, \citet{Wojtak13}, \citet{Velander14}, \citet{Han15}, and from the Milky Way. Note that the left-hand panels show the stellar mass at a given halo mass, $\langle M_*|M_\rmh\rangle$, while the right-hand panels show the halo mass at a given stellar mass, $\langle M_\rmh|M_*\rangle$.}
  \label{plot_GAMA_msmh}
\end{figure*}
\indent The constraints on the parameters are listed in Table \ref{tab_res}. The marginalised posteriors of the pairs of parameters are shown in Fig. \ref{plot_post}. This figure illustrates that the main degeneracies in the halo model occur between the parameters that describe the stellar-to-halo mass relation, and between the parameters that describe the CSMF of the satellites. These degeneracies are expected, given the functional forms that we adopted (see Eq. \ref{eq_msmh}, \ref{eq_phis}, \ref{eq:phi}). For example, a larger value for $M_{\rmh,1}$ would decrease the amplitude of the stellar-to-halo mass relation, which could be partly compensated by increasing $M_{*,0}$; hence these parameters are correlated. Similarly, increasing $b_0$ would lead to a higher normalisation of $\Phi_\rms(M_*|M)$, which could be partly compensated by decreasing $b_1$, hence these two parameters are anti-correlated. Furthermore, by comparing the marginalised posteriors to the priors, we observe that all parameters but one, $\beta_1$, are constrained by the data. We have verified that varying the prior on $\beta_1$ does not impact our results. Comparing the posteriors of the combined fit to the analysis where we only fit the lensing signal reveals that the stellar mass function helps by constraining several parameters; those that describe the stellar-to-halo mass relation and those that describe the satellite CSMF. \\
\indent In Fig. \ref{plot_GAMA_msmh}, we present the stellar-to-halo mass relation of central galaxies and the ratio of halo mass to stellar mass. The relation consists of two parts. This is not simply a consequence of adopting a double power law for this relation in the halo model, since the fit has the freedom to put the pivot mass below the minimum mass scale we probe, which would effectively result in fitting a single power law. For $M_*<5\times10^{10}h^{-2}M_\odot$, the stellar-to-halo mass relation is fairly steep and the stellar mass increases with halo mass as a powerlaw of $M_\rmh$ with an exponent $\sim$7.. At higher stellar masses, the relation flattens to $\sim$$M_\rmh^{0.25}$. The ratio of the dark matter to stellar mass has a minimum at a halo mass of $8\times10^{11}h^{-1}M_\odot$, where $M_*^\rmc=(1.45\pm0.32)\times10^{10}h^{-2}M_\odot$ and the halo mass to stellar mass ratio has a value of $M_\rmh/M_*=56_{-10}^{+16}$ [$h$]. The uncertainty on this ratio reflects the errors on our measurements and does not account for the uncertainty of the stellar mass estimates themselves, which are typically considerably smaller than the bin sizes we adopted and hence should not affect the results much. The location of the minimum is important for galaxy formation models, as it shows that the accumulation of stellar mass in galaxies is most efficient at this halo mass. \\
\indent In the lower left-hand panel of Fig. \ref{plot_GAMA_msmh}, we also show the integrated stellar mass content of satellite galaxies divided by halo mass. In haloes with masses $\gtrsim 2\times10^{13}h^{-1}M_\odot$, the total amount of stellar mass in satellites is larger than that in the central. At $5\times10^{14}h^{-1}M_\odot$ $\sim$94\% of the stellar mass is in satellite galaxies. Note that another considerable fraction of stellar mass  is contained in the diffuse intra-cluster light \citep[up to several tens of percents, see e.g.][]{Lin04} which we have not accounted for here. The recently discovered ultra-diffuse galaxies \citep[e.g.][]{Abraham14,Vanderburg16} form yet another source of unaccounted for stellar mass, but how much they contribute to the total stellar mass budget is currently uncertain. \\
\indent The normalisation of the mass-concentration relation is fairly low, $f_{\rm conc}=0.70^{+0.19}_{-0.15}$. A normalisation lower than unity was anticipated as we did not account for miscentring of centrals in the halo model. Miscentring distributes small-scale lensing power to larger scales, an effect similar to lowering the concentration. In our fits, it merely acts as a nuisance parameter, and should not be interpreted as conflicting with numerical simulations. In future work, we will include miscentering of centrals in the modelling, which should enable us to derive robust and physically meaningful constraints on $f_{\rm conc}$. The subhalo mass of satellites is not constrained by our measurements, which is why we do not show it in Fig. \ref{plot_post}. This is not surprising, given that most of our lenses are centrals, and that the lensing signal is fairly noisy at small projected distances from the lens. 

\subsection{Sensitivity tests on stellar-to-halo mass relation}\label{sec_sens}
\indent We have performed a number of tests to examine the robustness of our results. For computational reasons, we limited the number of model evaluations to 750\,000 (instead of \mbox{2\,100\,000}), divided over two chains. We adopted a maximum value of $R=1.05$ in the Gelman-Rubin convergence test to ensure that results are sufficiently robust to assess potential differences.  \\
\indent First, we test if incompleteness in our lens sample can bias the stellar-to-halo mass relation. As GAMA is a flux-limited survey, our lens samples miss the faint galaxies at a given stellar mass. If these galaxies have systematically different halo masses, our stellar-to-halo mass relation may be biased. To check whether this is the case, we selected a (nearly) volume-limited lens sample using the methodology of \citet{Lange15}. This method consists of determining a limiting redshift for galaxies in a narrow stellar mass bin, $z_{\rm lim}$, which is defined as the redshift for which at least 90\% of the galaxies in that sample have \mbox{$z_{\rm lim}<z_{\rm max}$}, with $z_{\rm max}$ the maximum redshift at which a galaxy can be observed given its rest-frame spectral energy distribution and given the survey magnitude limit. $z_{\rm lim}$ is determined iteratively using only galaxies with $z<z_{\rm lim}$. We removed all galaxies with redshifts larger than $z_{\rm lim}$ ($\sim$60\% of the galaxies in the first stellar mass bin, fewer for the higher mass bins) and repeated the lensing measurements. The resulting measurements are a bit noisier, but do not differ systematically. We fit our halo model to this lensing signal and the stellar mass function. The resulting stellar-to-halo mass relation becomes broader by up to 20\% at the low mass end, but is fully consistent with the result shown in Fig. \ref{plot_GAMA_msmh}. Hence we conclude that incompleteness of the lens sample is unlikely to significantly bias our results. \\
\indent We have also tested the impact of various assumptions in the set up of the halo model. We give details of these tests in Appendix \ref{app_sens}. None of the modifications led to significant differences in the stellar-to-halo mass relation, which shows that our results are insensitive to the particular assumptions in the halo model.

\subsection{Literature comparison}
\indent We limit the literature comparison to some of the most recent results, referring the reader to extensive comparisons between older works in \citet{Leauthaud12,Coupon15,Zu15}. Our main goal is to see whether our results are in general agreement. In-depth comparisons between results are generally difficult, due to differences in the analysis (e.g. the definition of mass, choices in the modelling) as well as in the data (e.g. the computation of stellar masses - note, however, that the stellar masses used in the literature stellar-to-halo mass relations we compare to are all based on a \citet{Chabrier03} IMF, as are ours).  \\
\indent \citet{Leauthaud12} measured the stellar-to-halo mass relation of central galaxies by simultaneously fitting the galaxy-galaxy lensing signal, the clustering signal and the stellar mass function of galaxies in COSMOS. The depth of this survey allowed them to measure this relation up to $z=1$. In Fig. \ref{plot_GAMA_msmh}, we show their relation for their low-redshift sample at 0.22$<$$z$$<$0.48, which is closest to our redshift range. The relations agree reasonably well. We infer a slightly larger halo mass at a given stellar mass, most noticeably at the high mass end. Their $M_\rmh/M_*$ ratio reaches a minimum at a halo mass of $8.6\times10^{11}h^{-1}M_\odot$ with a value of \mbox{$M_\rmh/M_* =38$ [$h$]}, $\sim$1.5$\sigma$ below the minimum value of the ratio we find. A systematic shift in stellar mass may explain much of the difference. The stellar masses used in \citet{Leauthaud12} are based on photometric redshifts, which generally induce a small Eddington bias in the stellar mass estimates, particularly at the high-stellar mass end where the stellar mass function drops exponentially \citep[as illustrated in Fig. 4 in][]{Drory09}. In contrast, the stellar masses in GAMA are computed using spectroscopic redshifts, and this bias does not occur. We attempted various shifts in stellar mass and found that the stellar-to-halo mass relations fully overlap if we decrease the stellar masses from \citet{Leauthaud11} by 0.15 dex. Systematic differences between stellar mass estimates from the literature are typically of this order \citep[see e.g.][]{Mobasher15}, which implies that the accuracy of the stellar-to-halo mass relation is already limited by systematic uncertainties in the stellar mass estimates. \\
\indent Next, we compare our results to \citet{Moster13}, who applied an abundance matching technique to the Millennium simulation \citep{Springel05}. The stellar mass functions were adopted from various observational studies, but were all converted to agree with a \citet{Chabrier03} IMF. We use the fitting functions provided in that work to compute the stellar-to-halo mass relation at $z=0.25$, close to the mean redshift of our full sample. We find good agreement between the results as shown in Fig. \ref{plot_GAMA_msmh}. The minimum of their $M_\rmh/M_*$ ratio is located at $7\times10^{11}h^{-1}M_\odot$, close to our best-fit result of $8\times10^{11}h^{-1}M_\odot$. At this location, their halo to stellar mass ratio takes a value of \mbox{$M_\rmh/M_* =50$ [$h$]}. \\
\indent We compare our measurements to the results of \citet{Han15} in the right-hand panel of Fig. \ref{plot_GAMA_msmh}. \citet{Han15} measured halo masses for the same GAMA sample, but using sources from the SDSS. Halo masses were estimated for a volume-limited lens sample using a maximum likelihood technique. In contrast to our work, their measurements show the average halo mass for a given stellar mass, which is not the same due to the intrinsic scatter \citep[see e.g. Fig. (7) of][]{Tinker13}. Hence we converted our results using Bayes theorem \citep[see e.g.][]{Coupon15} to enable a comparison. We find excellent agreement between the results. \\
\indent \citet{Wojtak13} present halo mass estimates for galaxies in stellar mass bins obtained from the kinematics of satellite galaxies around isolated galaxies in the SDSS. Halo masses were defined with respect to $\rho_{\rm crit}$ instead of the mean density, which are typically 30-40\% smaller. To account for this, we multiplied their masses with a factor 1.3. We find good agreement at stellar masses $M_*<8\times 10^{10}h^{-2} M_\odot$, but at higher stellar masses their halo masses are somewhat lower than ours. A potential reason is that their sample only consists of isolated galaxies, which may have systematically lower halo masses. Also note that they remark in their work that their halo masses are $\sim$0.2 dex lower at the high-mass end than what is typically reported in the literature. \\
\indent Finally, we compare our results to the galaxy-galaxy lensing results from \citet{Velander14}, who measured the lensing signal around red and blue galaxies at 0.2$<$$z$$<$0.4 in CFHTLenS over a large range in stellar mass. Masses were defined with respect to $\rho_{\rm crit}$, which we multiplied with a factor 1.3 to convert them to our definition. The agreement is fair; we find a good match at low stellar masses, but for $M_*>5\times 10^{10}h^{-2} M_\odot$, our halo masses are somewhat larger. What may be contributing to this difference, is that \citet{Velander14} inferred a relatively high satellite fraction for red galaxies at the high stellar mass end (reaching as high as their upper limit of 0.2), which may have pushed their average halo mass down. Also, as for \citet{Leauthaud12}, stellar masses were determined using photometric redshifts, which may have induced a small Eddington bias. If we decrease their mean stellar masses by 0.1 dex, their measurements fully overlap with ours. \\
\indent Summarising the above, we conclude that although we find small differences between our stellar-to-halo mass relation and those from the literature, the agreement is fair in general. 

\subsubsection{Milky Way comparison}
\begin{figure*}
  \resizebox{\hsize}{!}{\includegraphics{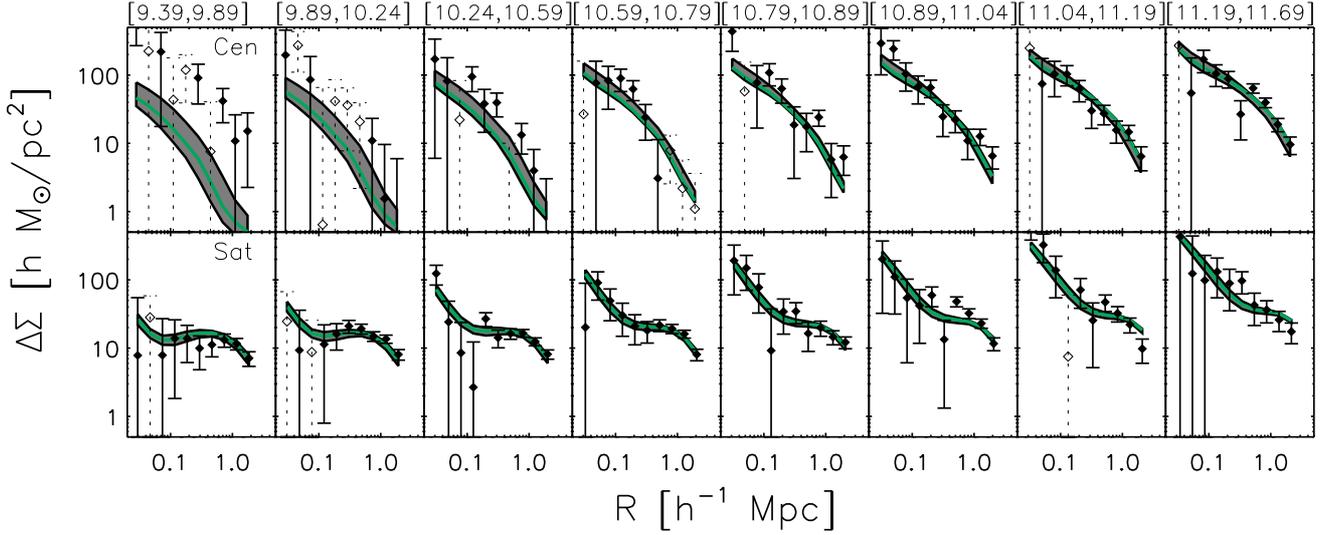}}
  \caption{Excess surface mass density profile of GAMA galaxies measured as a function of projected separation to the lens, selected in various stellar mass bins as indicated at the top of each column, that are centrals (top row) and satellites (bottom row) of `rich' ($N_{\rm fof}\ge 5$) groups. The bin ranges correspond to the $\log_{10}$ of the stellar mass and are in units $\log_{10}(h^{-2}M_\odot)$. Open symbols and the dashed lines indicate the absolute value of the negative data points and their errors. The green solid line indicates the best-fit halo model, the grey contours indicate the 68\% model uncertainty.}
  \label{plot_GAMA_gg_all_envir}
\end{figure*}
\begin{figure}
  \resizebox{\hsize}{!}{\includegraphics{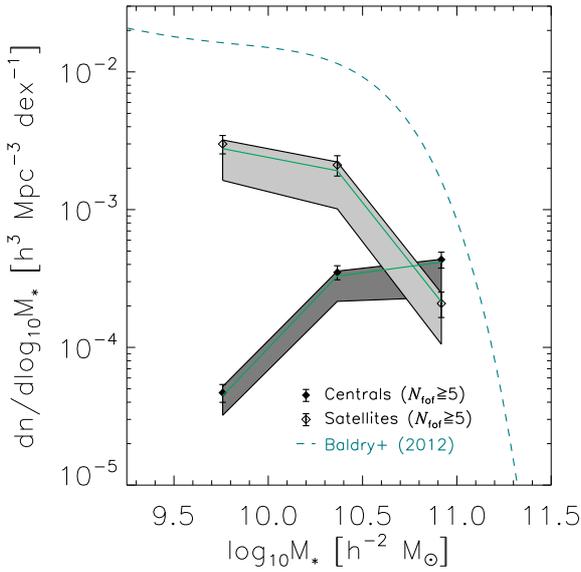}}
  \caption{Stellar mass functions for GAMA galaxies at $z<0.15$ that are centrals and satellites in `rich' groups, for a comoving volume. Errors have been determined by bootstrap and include the contributions from Poisson noise and cosmic variance. The green solid line indicates the best-fit halo model, the grey regions indicate the 68\% model uncertainty, linearly interpolated between the stellar mass bins. We also show the analytical fit to the SMF from \citet{Baldry12} for all galaxies for reference. The model uncertainties are somewhat skewed with respect to the data and the best-fit model, which is caused by sampling issues, as discussed in the text.}
  \label{plot_GAMA_smf_envir}
\end{figure}

\indent The number of satellite galaxies depends on halo mass, and observations of the Milky Way suggest that it may have fewer satellites than expected given its stellar mass \citep[e.g.][]{Klypin99,Moore99}. To resolve this so-called ‘missing satellite problem’ various studies have shown that the tension is eased for lower halo masses of the Milky Way \citep[e.g.][]{Wang12,Vera13}. This raises the question whether or not the location of the Milky Way is special in the stellar mass to halo mass plane, i.e. whether its halo mass is peculiarly low given its stellar mass. \\
\indent Total stellar mass estimates for the Milky Way are typically of order $(6\pm1)\times10^{10}M_\odot$ \citep{McMillan11,Licquia15}. Halo mass estimates have a considerably larger scatter, with $M_{200}$ estimates ranging between $(0.5-2) \times 10^{12} M_\odot$ \citep[see Fig. 1 of][]{Wang15}. We adjust these local measurements assuming $h=0.7$ and show the results in Fig. \ref{plot_GAMA_msmh}. Our stellar-to-halo mass relation predicts a mean stellar mass of $1.8\times10^{10}h^{-2}M_\odot$ at a halo mass of $1\times10^{12}h^{-1}M_\odot$. The Milky Way lies just at the edge of our 1$\sigma$ contours. However, our confidence intervals only correspond to the uncertainties on the mean relation, and when comparing individual objects, one should take the intrinsic scatter between stellar and halo mass, which is $\sim$0.2 dex, into account. Hence the lower limit on the stellar mass of the Milky Way ($5\times10^{10}M_\odot$) is roughly 1$\sigma$ away in terms of intrinsic scatter at $M_\rmh=10^{12}h^{-1}M_\odot$. Although the Milky Way appears to have a relatively high stellar mass given its halo mass, it is not particularly anomalous.

\subsection{Satellite fraction}\label{sec_satfrac}
Our sample consists of a mixture of central and satellite galaxies. In the halo model we fit for the contribution of both, which enables us to determine the satellite fraction using Eq. (\ref{eq_fsat}). The results are shown in Fig. \ref{plot_fsat}. The satellite fraction decreases with stellar mass from $\sim$0.3 at $5\times10^9 h^{-2}M_\odot$ to $\sim$0.05 at $2\times10^{11} h^{-2}M_\odot$. Particularly at the high-mass end, it is well constrained. As for the stellar-to-halo mass relation, including the constraints from the stellar mass function has a significant impact and considerably decreases the model uncertainty. The satellite fraction does not sensitively depend on assumptions in the halo model, as discussed in Appendix \ref{app_sens}.  \\
\indent We compare our satellite fractions to those based on the GAMA group catalogue. For every stellar mass range, we count all galaxies listed as satellite (not restricted to groups with $N_{\rm fof}\ge5)$, and divide that by the total number of galaxies in that range. We only include GAMA galaxies at $z<0.3$ here, to reduce the impact of incompleteness. The resulting satellite fractions do not sensitively depend on the specific value of the redshift cut. The ratio is shown as the upper dashed line in Fig. \ref{plot_fsat}. It provides an estimate of the true satellite fraction, but a crude one as the group membership identification in GAMA becomes less robust towards groups with fewer members \citep{Robotham11} and we do not apply a cut on $N_{\rm fof}$. Hence a fraction of the galaxies that are labelled as satellites may in fact be centrals. In addition, some satellites may not be identified and as such be excluded from the group catalogue. We derive a more robust lower limit on the satellite fraction by only counting the satellites in `rich' groups ($N_{\rm fof} \ge5$) and dividing that by the total number of galaxies in that stellar mass range. This is indicated by the lower dotted line in Fig. \ref{plot_fsat}. The satellite fraction we obtain from the halo model should be larger than this, which we find to be the case at $M_*<10^{11} h^{-2}M_\odot$. For higher stellar masses, our constraints on the satellite fraction fall below the lower limit from the GAMA catalogue. Albeit not very significant, it suggests that a fraction of satellites at the high stellar mass end are actually centrals, or that one or more assumptions in our halo model are inaccurate. Either way, it shows that the combination of galaxy-galaxy lensing with the stellar mass function has the potential to become a valuable tool to infer the robustness of group catalogues. We expect that including the clustering of galaxies in the fit will further tighten the constraints on the satellite fraction \citep[see e.g.][]{Cacciato09}.

\section{Environmental dependence}\label{sec_envir}
\indent Galaxies in groups are subject to processes such as quenching, stripping and merging. One of the observable consequences is that star formation is suppressed and galaxies turn red \citep[see e.g.][]{Boselli06}. The cumulative impact of these processes is likely to affect the baryonic and dark matter content of centrals and satellites in different ways. An infalling (satellite) galaxy, for example, is expected to lose relatively more dark matter than stars, as the latter mainly reside in the central part of the halo where the potential well is deep \citep[e.g.][]{Wetzel14}. This increases the group halo mass, but should not affect the stellar mass of the central much. If, on the other hand, an already accumulated satellite that has been stripped off its dark matter merges with the central galaxy, the stellar mass of the central increases, but not the halo mass. By comparing the stellar-to-halo mass relation for centrals in `rich' groups to the one of the full sample, we can study the relative importance of such environmental effects. \\

\subsection{Centrals in rich groups}
\begin{figure}
  \resizebox{\hsize}{!}{\includegraphics{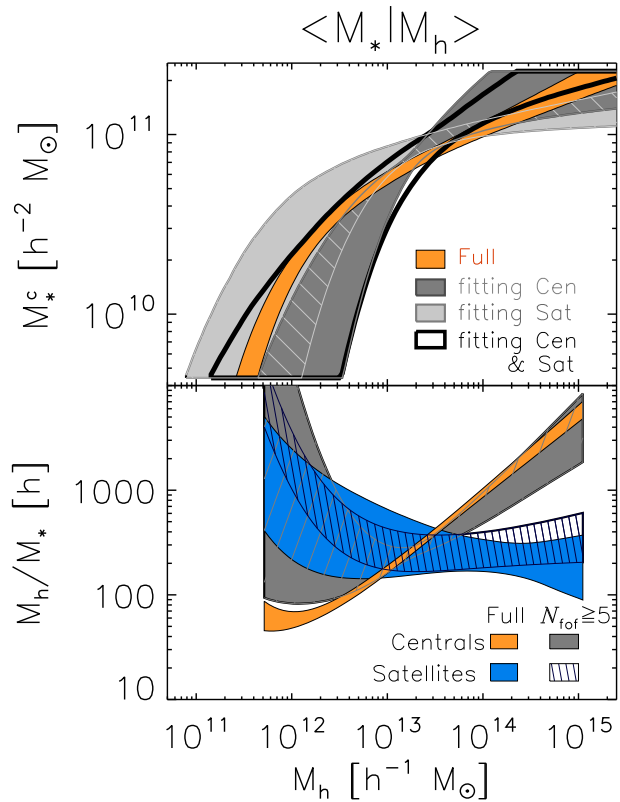}}
  \caption{68\% confidence intervals of the stellar-to-halo mass relation for central galaxies in `rich' groups ($N_{\rm fof}\ge5$), determined using the lensing signal and stellar mass function of central galaxies (`Cen', dark grey), of satellites galaxies (`Sat', light grey) and of all galaxies in `rich' groups (`Cen \& Sat', empty contours). For reference, we also show the relation from the full sample in orange. The bottom panel shows the halo mass to stellar mass ratio of the centrals and satellites in `rich' groups (dark grey and dark blue hatched, respectively). The halo mass to stellar mass ratio of the full sample is shown for reference (orange and light blue for centrals and satellites, respectively). }
  \label{plot_GAMA_msmhl_envir}
\end{figure}

We first select the central galaxies in `rich' groups (with a multiplicity $N_{\rm fof}\geq5$) and measure their lensing signal and stellar mass function using the same binning as before. We exclude groups with fewer than five members because comparisons with mock data have shown that those are affected more by interlopers \citep{Robotham11}, which makes an interpretation of the results harder. The galaxy-galaxy lensing signal is shown in Fig. \ref{plot_GAMA_gg_all_envir} and the stellar mass function in Fig. \ref{plot_GAMA_smf_envir}. We measure the stellar mass function using groups at $z<0.15$ to ensure the sample is volume-limited. \\
\indent To fit the halo model to the data, we have to account for one additional complication. For certain sub-samples of galaxies, not all haloes of mass $M$ contain a central galaxy, whilst the halo model assumes that all of them do (the integral of Eq. \ref{eq:phi_c} over stellar mass is unity). We account for this by introducing a `halo mass incompleteness' factor, a generic function that varies between 0 and 1, which we multiply with the CSMF of the central galaxies:
\begin{equation}
 {\widetilde\Phi_\rmc(M_*|M_\rmh)} =  \Phi_\rmc(M_*|M_\rmh) \times \textrm{erf} (c_0 [\log_{10}(M_\rmh)-c_1]),
 \label{eq_incompl}
\end{equation}
with $c_0$ and $c_1$ two incompleteness parameters that we fit for; $c_1$ determines where we the transition to incompleteness occurs, and $c_0$ determines how smooth or abrupt the transition is. This incompleteness factor is only suitable for selections of lenses whose abundance as a function of stellar mass increases/decreases monotonically with respect to the full sample, as is the case here. A similar approach was taken in \citet{Tinker13} in order to simultaneously measure the stellar-to-halo mass relation of quiescent and star-forming galaxies  (see their Sect. 3.2). \\
\indent To fit the halo model, we need to apply priors on the incompleteness parameters $(c_0,c_1)$, as the large covariance of the three stellar mass function bins results in a peculiar likelihood surface. For a large range of $(c_0,c_1)$ values, the  $\chi^2$ of the stellar mass function is high but practically constant. When the MCMC chains start far from the minimum, they can get stuck in this $\chi^2$ plateau. To avoid having to run very long chains to ensure all walkers find their way to the minimum, we adopt flat priors and restrict $c_0$ to $[-5,5]$ and $c_1$ to $[9,14]$ which generously brackets the best-fit for any chain we run and hence should not affect the results. Note that we adopt a range of $[9,16]$ for $c_1$ for all other runs, see Table \ref{tab_prior}. We start the chains close to the best-fit location, as determined from a previous run, to avoid that many walkers start in this $\chi^2$ plateau and never reach the minimum. Even with these precautions, a fraction (10\%) of the chains remain stuck\footnote{ We also implemented a Metropolis-Hastings sampler with a proposal distribution derived by a Fisher information matrix analysis and found it also suffered from sampling problems which could not be solved by adjusting the step-sizes.}. Since those models have a similar (large) $\chi^2$ contribution from the stellar mass function, we can easily identify them and remove them before we analyse the chain. Figure \ref{plot_GAMA_smf_envir} shows that the model uncertainty of the stellar mass function is somewhat skewed with respect to the data and the best-fit halo model. The reason is that some of the walkers are close to the $\chi^2$ plateau and still in the process of evolving towards the minimum. We have checked that including these problematic walkers does not affect our results. \\
\indent The best-fit model has a reduced $\chi^2$ of $98/(83 - 8)=1.3$. Figure \ref{plot_GAMA_gg_all_envir} shows that the lensing signal of the lowest stellar mass bin is not well fit. A possible reason is that the lowest stellar mass samples are contaminated with satellite galaxies, for which we provide evidence in Sect. \ref{sec_satfrac_envir}. Note that the lensing signal of these bins are very noisy and that a potential bias of the stellar-to-halo mass relation at the low-mass end resulting from this contamination is unlikely to be significant. \\
\indent The constraints on the fit parameters are tabulated in Table \ref{tab_res}. The stellar-to-halo mass relation is shown in Fig. \ref{plot_GAMA_msmhl_envir}. The 68\% confidence interval is broader than the one of the full sample due to the noisier lensing measurements at the low stellar mass end. Nonetheless, it shows that the stellar-to-halo mass relation of centrals in `rich' groups is consistent with the relation for the full sample, suggesting that this relation does not sensitively depend on local density. Note that centrals in `rich' groups form $\sim$15\% of the total lens sample for the two highest stellar mass bins, so the average lensing signals of centrals in those bins are somewhat correlated to the lensing signals of the corresponding bins of the full sample (and consequently, the stellar-to-halo mass relations will be correlated as well at the high-mass end).  \\
\indent We also inferred the stellar-to-halo mass relation from the lensing signal only. The resulting minimum $\chi^2$ is 97 with 72 degrees of freedom, hence a similar reduced $\chi^2$ as for the combined fit. The stellar-to-halo mass relation is consistent, but less well constrained than the combined fit, particularly at stellar masses $>2\times10^{10}h^{-2}M_\odot$, where the upper limit in halo masses is shifted to larger values, already extending to $10^{15}h^{-1}M_\odot$ at stellar masses of $8\times10^{10}h^{-2}M_\odot$. \\
\indent Our result appears somewhat at odds with \citet{Tonnesen15}, who studied environmental variations of the stellar-to-halo mass ratio using a large suite of cosmological hydrodynamical simulations.  Environments were classified according to the mean density on 20 Mpc scales, stellar mass were computed by adding the mass of all star particles that belonged to a galaxy. They reported a significantly larger stellar-to-halo mass ratio for galaxies in large-scale overdensities, compared to those in large-scale underdensities. Their most massive halo mass bin extends to $10^{13}h^{-1}M_\odot$. In this regime, we find that the stellar-to-halo mass ratio of centrals is not larger than average. Note, however, that it is difficult to compare the results, as the samples were selected in very different ways. The difference in local density is smaller in our work (comparing central galaxies as tracer of overdense regions, to the average of all environments). Also, if the stellar masses of \citet{Tonnesen15} systematically include more stellar mass from the outskirts of galaxies, this may partly explain the difference between the results. We plan to perform a more direct comparison in a future work, where we will measure the stellar-to-halo mass relation in knots, filaments, sheets and voids, using the environment catalogues from \citet{Alpaslan14} and \citet{Eardley15}, as well as according to local density estimates as employed in \citet{Tonnesen15}. \\

\subsection{Satellites in rich groups}
Next, we analysed the galaxies listed as satellites in `rich' groups ($N_{\rm fof}\geq5$) performing a simultaneous fit to both the lensing signal and the stellar mass function. Similar to our analysis of the centrals in the previous section, we account for incompleteness by multiplying the CSMF of the satellites by the `halo mass incompleteness' factor,
\begin{equation}
 {\widetilde\Phi_\rms(M_*|M_\rmh)} =  \Phi_\rms(M_*|M_\rmh) \times \textrm{erf} (c_0 [\log_{10}(M_\rmh)-c_1]).
 \label{eq_incompl2}
\end{equation}
The lensing signals in bins of stellar mass and the stellar mass function are shown in Fig. \ref{plot_GAMA_gg_all_envir} and \ref{plot_GAMA_smf_envir}. In the fit, we fixed the stellar-to-halo mass relation of the centrals, as well as the incompleteness parameters in Eq. (\ref{eq_incompl}), to the best-fit values of our nominal results for centrals from the previous section. Also here the large covariance of the stellar mass function resulted in a broad likelihood surface: when fitting the halo model with the standard priors (listed in Table \ref{tab_prior}), part of the chain would get stuck in regimes far from the minimum (at $\Delta \chi^2 \sim 20$). To avoid this, we excluded the part of parameter space where this problem occurred through the prior $b_0>-0.5$. We ensured that this did not exclude the regime close to the minimum. With this additional prior, the minimisation ran smoothly. The resulting best-fit model has a reduced $\chi^2$ of $74/(83 - 5)=0.9$. The constraints on the satellite CSMF parameters, and on $f_{\rm conc}$ and $f_{\rm sub}$ (the normalisation of the mass-concentration relation and the subhalo mass fraction) are listed in Table \ref{tab_res}. Since the confidence interval of $b_0$ is sensitive to where we put this prior, it should be interpreted with care. \\
\indent The concentration of the subhaloes is consistent with predictions from dark-matter-only simulations. Since the sample now only consists of satellites, we derive much tighter constraints on the subhalo mass fraction: \mbox{$f_{\rm sub}=0.25^{+0.09}_{-0.08}$}. This result is robust against changes in the halo model, as detailed in Appendix \ref{app_sens}. \\
\indent The subhalo mass fraction was also determined in \citet{Sifon15} for the same satellites, but now separated into samples at different distances from their hosts. They reported a subhalo mass fraction in the range 1-2\%. When we fix the subhalo mass fraction in our halo model to such low values, the $\chi^2$ values of the fit significantly degrades, and the model underestimates the lensing signal at small scales for the high stellar mass bins. We attribute this difference to the fact that \citet{Sifon15} average over all stellar masses. Figure \ref{plot_GAMA_gg_all_envir} shows that the small-scale lensing signal of the first three stellar mass bins is very small and noisy. These two bins contain 2/3 of all lenses stacked in \citet{Sifon15}, and pull the average lensing signal down. When we separate the lensing signal in stellar mass bins, most of the constraining power comes from the massive stellar mass bins, which have the highest lensing signal-to-noise ratio. These bins clearly prefer a larger subhalo mass fraction. The subhalo mass fraction is not pulled down by the lower mass bins, as their signal is noisy and can accommodate a higher subhalo mass fraction. \\
\indent These results suggest a subhalo halo mass fraction that increases with host mass. To test this, we parametrised the subhalo mass fraction as $A_{\rm sub} \times (\langle M_*/h^{-2} M_\odot\rangle /10^{10.5})^{\alpha_{\rm sub}}$ and fit for $A_{\rm sub}$ and $\alpha_{\rm sub}$. The fit slightly improves with a minimum $\chi^2$ of 71, and we obtain $A_{\rm sub}=0.15^{+0.12}_{-0.08} h^2 M_\odot^{-1}$ and $\alpha_{\rm sub}=0.34^{+0.36}_{-0.32}$, providing weak evidence that the subhalo mass fraction increases with stellar mass. Such a trend would be supported in a scenario where the most massive subhaloes were accreted most recently and had not much time to be stripped of their dark matter \citep[see e.g.][]{RodriguezPuebla12}. Note, however, that an increasing contamination of central galaxies in the satellite sample towards higher stellar masses, would be able to mimic such a trend as well. In fact, in Sect. \ref{sec_satfrac_envir} we measure the satellite fraction and find evidence for such a contamination. When we let the satellite fraction free in the fit,  $f_{\rm sub}$ favours smaller values ($f_{\rm sub}=0.06^{+0.13}_{-0.04}$).  \\ 
\indent The lensing signal of the satellites indirectly constrains the stellar-to-halo mass relation of the centrals of the haloes that host them. On small scales, the lensing signal is determined by a combination of the host halo mass and the subhalo mass fraction (assumed to be constant again), and on scales of $\sim$1 Mpc, the hosting haloes cause the characteristic bump in the lensing signal of the satellites, whose amplitude depends on the average halo mass \citep[see Sec. 3 of][]{Sifon15}. If a mixed sample of centrals and satellites is used to measure the stellar-to-halo mass relation, one can induce a bias if the satellite contribution is not properly modelled. Here, we have the data in hand to test whether this is the case in our modelling. Hence we perform a halo model fit where we additionally fit for the five parameters that describe the stellar-to-halo mass relation of the centrals and the two incompleteness parameters, using the measurements of the satellites only. With this set-up, the halo model provides satisfactory fits, with a best fit reduced $\chi^2$ of $62/(83 - 12)=0.9$. The parameter constraints are listed in Table \ref{tab_res}. \\
\indent In Fig. \ref{plot_GAMA_msmhl_envir}, we show the constraints on the stellar-to-halo mass relation of central galaxies, obtained by fitting the lensing signal and stellar mass function of the satellites only. The satellites prefer a somewhat steeper relation at the high mass end. The contours are mostly overlapping with those from the fit to the centrals. This is an important test of the halo model, as it shows that the uncertainties on the assumptions in modelling the satellite signal do not lead to large biases in the stellar-to-halo mass relation when we fit a mixed sample of centrals and satellites. For comparison, we also show the stellar-to-halo mass relation of centrals when we fit all the galaxies in `rich' groups simultaneously (hence fitting for the satellite fraction). The results are consistent with the fits to the centrals/satellites only. \\
\indent It is interesting to note that the uncertainty on the stellar-to-halo mass relation of centrals is actually smaller at \mbox{$M_*>10^{11}h^{-2}M_\odot$} when measured from the satellite signal. This counter-intuitive result can be understood as follows: groups contain more satellites than centrals. Stacking many satellites reduces the statistical uncertainties, which is somewhat counteracted by an increased correlation between the radial bins of the lensing signal as the same background galaxies are used multiple times. Furthermore, haloes with more satellites, which are typically more massive, get a larger effective weight. Moreover, we obtain additional constraints on the parent halo mass from the lensing signal on small scales, as we fit the subhalo mass fraction as a constant (which is an implicit prior).\\
\indent Finally, we also determine the total stellar mass content in satellites at a given group halo mass, by integrating over the CSMF of the satellites. The results are shown in Fig. \ref{plot_GAMA_msmhl_envir}. The constraints are tighter than those of the full sample, but completely consistent.

\subsection{Satellite fractions}\label{sec_satfrac_envir}
\begin{figure}
  \resizebox{\hsize}{!}{\includegraphics{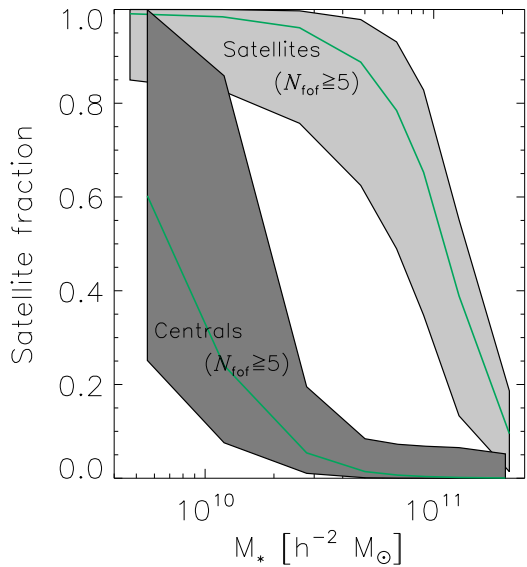}}
  \caption{Halo model test: 68\% model uncertainty on the fitted satellite fractions for centrals and satellites in `rich' groups, as indicated in the figure. The best-fit models are indicated by the green solid lines.}
  \label{plot_GAMA_fsat_envir}
\end{figure}
\indent In our nominal runs, we fix the satellite fractions to zero and unity when we analyse central and satellite samples, respectively. The identification of centrals and satellites is, however, only robust in groups with a multiplicity $N_{\rm fof}\geq5$ \citep{Robotham11}. This opens up the possibility to perform a unique halo model test: we can run it without informing the halo model of the nature of the lens sample. The resulting satellite fraction constraints should be consistent with 0 and 1 for the centrals and satellites, respectively. A failure would indicate a problem with the halo model, or point at impurities in the group catalogue. When we carry out this test on the satellite sample, some parts of the chain get stuck at the aforementioned $\chi^2$ plateau. As these regimes had clearly distinct $\chi^2$ contributions from the stellar mass function, they could easily be identified and removed before analysing the chains. \\
\indent The satellite fractions are shown in Fig. \ref{plot_GAMA_fsat_envir}. For the centrals, we find that the satellite fraction is consistent with zero, although the uncertainty becomes very large at the low stellar mass end, where the lensing signals are noisy. At the high stellar mass end, we derive an upper limit on the satellite fraction of 0.05 at 1$\sigma$. Varying assumptions in the halo model only changes the size of the confidence interval, as discussed in Appendix \ref{app_sens}.  \\
\indent For the satellite sample, the satellite fraction is consistent with unity at $M_*<10^{11}h^{-2} M_\odot$, with a lower limit of 0.85. At higher stellar masses, the satellite fraction drops and has a value of 0.10 for our most massive bin. To test if this is the result of particular choices in our halo model, we vary the list of assumptions from Appendix \ref{app_sens} and determine the satellite fraction in each run. In all cases, the resulting satellite fractions remain consistent with unity at $M_*<10^{11}h^{-2} M_\odot$, but drops at higher masses. How quickly it decreases, and the size of the uncertainties, depends on the halo model set up (see Appendix \ref{app_sens}). \\
\indent These results suggest that a substantial fraction of satellites with large stellar masses in `rich' groups, are in fact centrals that reside at the centre of the halo. The absence of a corresponding increase of the satellite fraction of the central sample, suggests that the misidentification of satellites could be due to groups actually consisting of the projection of two or more groups along the line-of-sight, or of being in the process of a merger. Alternatively, it could mean that the most massive satellite galaxies are not following the NFW profile of the dark matter, but are more centrally concentrated, residing closer to the centre of the group. Note that the number of affected objects is very small, as there are only few satellites galaxies with stellar mass $M_*>10^{11}h^{-2} M_\odot$; the group catalogue overall remains very pure.\\ 
\indent This test shows that the constraints from galaxy-galaxy lensing and the stellar mass function can be used to test the performance of group finders. Future, higher signal-to-noise data sets, combined with clustering data, will be able to constrain the satellite fraction at the few per cent level. These datasets therefore form a valuable complement for demonstrating the fidelity of group catalogues.

\section{Conclusions}\label{sec_conc}
In this work, we have studied how galaxies are related to their dark matter haloes by measuring their stellar-to-halo mass relation, and whether this relation depends on environment. We used data from the $\sim$100 deg$^2$ overlap between the GAMA and KiDS surveys: the former provides the information about intrinsic lens properties, as well as the group catalogue which enabled us to select galaxies in groups (dense environments), whilst we used the shape measurements and photometric redshift catalogues from KiDS to measure the lensing signal around the GAMA galaxies. \\
\indent The stellar-to-halo mass relation of central galaxies is poorly constrained from the lensing signal alone, the reason being that in the halo model predictions of the weak lensing signal around galaxies, lower halo masses can be partially compensated by higher satellite fractions (as satellites typically reside in more massive haloes). Thus informative priors need to be adopted on the satellite fraction to constrain the stellar-to-halo mass relation from lensing alone. This can be avoided by including the stellar mass function, which provides sufficient additional constraints to break this degeneracy: both the stellar-to-halo mass relation and the satellite fraction are better constrained when the lensing signal and the stellar mass function are fitted simultaneously. \\
\indent The stellar-to-halo mass relation can be described by a double power law. At the high mass end ($M_*>5\times10^{10}h^{-2}M_\odot$), the stellar mass increases with halo mass as $\sim$$M_\rmh^{0.25}$. The ratio of the dark matter to stellar mass has a minimum at a halo mass of $8\times10^{11}h^{-1}M_\odot$ with a value of $M_\rmh/M_*=56_{-10}^{+16}$ [$h$]. Our constraints are in fair agreement with recent results from the literature, although small, systematic shifts in stellar mass (of order 0.10-0.15 dex) can improve the agreement. Systematic differences between different stellar masses estimates \citep[due to different assumptions in the SED modelling, see e.g.][]{Coupon15} are typically of this order and hence already form a limiting factor in comparisons of stellar-to-halo mass relations from different works. This illustrates the need for reducing systematic errors in stellar mass estimates (which could be achieved by adopting standardized stellar mass measures). \\
\indent For the first time, we determined the stellar-to-halo mass relation of centrals in dense environments. We made use of the GAMA group catalogue to select galaxies that reside in `rich' groups (with a multiplicity $N_{\rm fof}\ge5$). We analysed the signals of both central galaxies and satellite galaxies separately. We fit the halo model in an informed setting, exploiting our prior knowledge of whether the sample contained centrals or satellites. The stellar-to-halo mass relation of central galaxies, determined from fitting the signals of the centrals, was consistent with the one determined from fitting the signal of the satellites, providing evidence that the uncertainties of the assumptions in modelling the satellite contribution in the halo model does not lead to biases when a mixed sample of centrals and satellites is used to measure the stellar-to-halo mass relation. \\
\indent  Interestingly, we find no large differences between the stellar-to-halo mass relation from all galaxies, and from those that reside in `rich' groups. This shows that the stellar-to-halo mass relation depends only weakly on environment. \\
\indent The group catalogue enables another unique test of the halo model: we fitted the signals of the centrals/satellites in `rich' groups, but without a fixed satellite fraction. The recovered satellite fractions are consistent with 0 for the centrals. For the satellites, we find an indication for an impurity in the group catalogue at the high stellar mass end. This shows that galaxy-galaxy lensing, combined with the stellar mass function (and in the future also clustering), can be used as important robustness tests for the correct identification of centrals/satellites in group finding algorithms. \\
\indent The average subhalo masses of satellites in `rich' groups are typically 25\% of their host haloes. These constraints are driven by massive satellites, which have the highest lensing signals. We find weak evidence for a subhalo mass fraction that increases with stellar mass, which would be consistent with the scenario where the most massive satellites are accreted most recently and still retain most of their dark matter. We cannot, however, draw definite conclusions as impurities in the satellite sample, as is implied by the test described above, could mimic such a trend. \\

\paragraph*{Acknowledgements}
We thank the referee for his/her comments that improved the draft. EvU acknowledges support from a grant from the German Space Agency DLR and from an STFC Ernest Rutherford Research Grant, grant reference ST/L00285X/1. MC, HHo, CS and MV acknowledge support from the European Research Council under FP7 grant number 279396. MV additionally acknowledges support from the Netherlands Organisation for Scientific Research (NWO) through grants 614.001.103. AC and CH acknowledge support from the European Research Council under the FP7 grant numbers 240185 and 647112. HHi is supported by an Emmy Noether grant (No. Hi 1495/2-1) of the Deutsche Forschungsgemeinschaft. BJ acknowledges support by an STFC Ernest Rutherford Fellowship, grant reference ST/J004421/1. This work is supported by the Deutsche Forschungsgemeinschaft in the framework of the TR33 `The Dark Universe'. This work is based on data products from observations made with ESO Telescopes at the La Silla Paranal Observatory under programme IDs 177.A-3016, 177.A-3017 and 177.A-3018. GAMA is a joint European-Australasian project based around a spectroscopic campaign using the Anglo-Australian Telescope. The GAMA input catalogue is based on data taken from the Sloan Digital Sky Survey and the UKIRT Infrared Deep Sky Survey. Complementary imaging of the GAMA regions is being obtained by a number of independent survey programs including GALEX MIS, VST KiDS, VISTA VIKING, WISE, Herschel-ATLAS, GMRT and ASKAP providing UV to radio coverage. GAMA is funded by the STFC (UK), the ARC (Australia), the AAO, and the participating institutions. The GAMA website is http://www.gama-survey.org/. \\ 
\indent {\it Author Contributions}: All authors contributed to the development and writing of this paper. The authorship list reflects the lead authors
(EvU, MC, HHo) followed by two alphabetical groups. The first alphabetical group includes those who are key contributors to both the scientific analysis and the data products. The second group covers those who have either made a significant contribution to the data products, or to the scientific analysis.
\bibliographystyle{mnras}

\begin{appendix}
\section{Covariance of stellar mass function}\label{app_smfcovar}
\begin{figure*}
  \resizebox{\hsize}{!}{\includegraphics{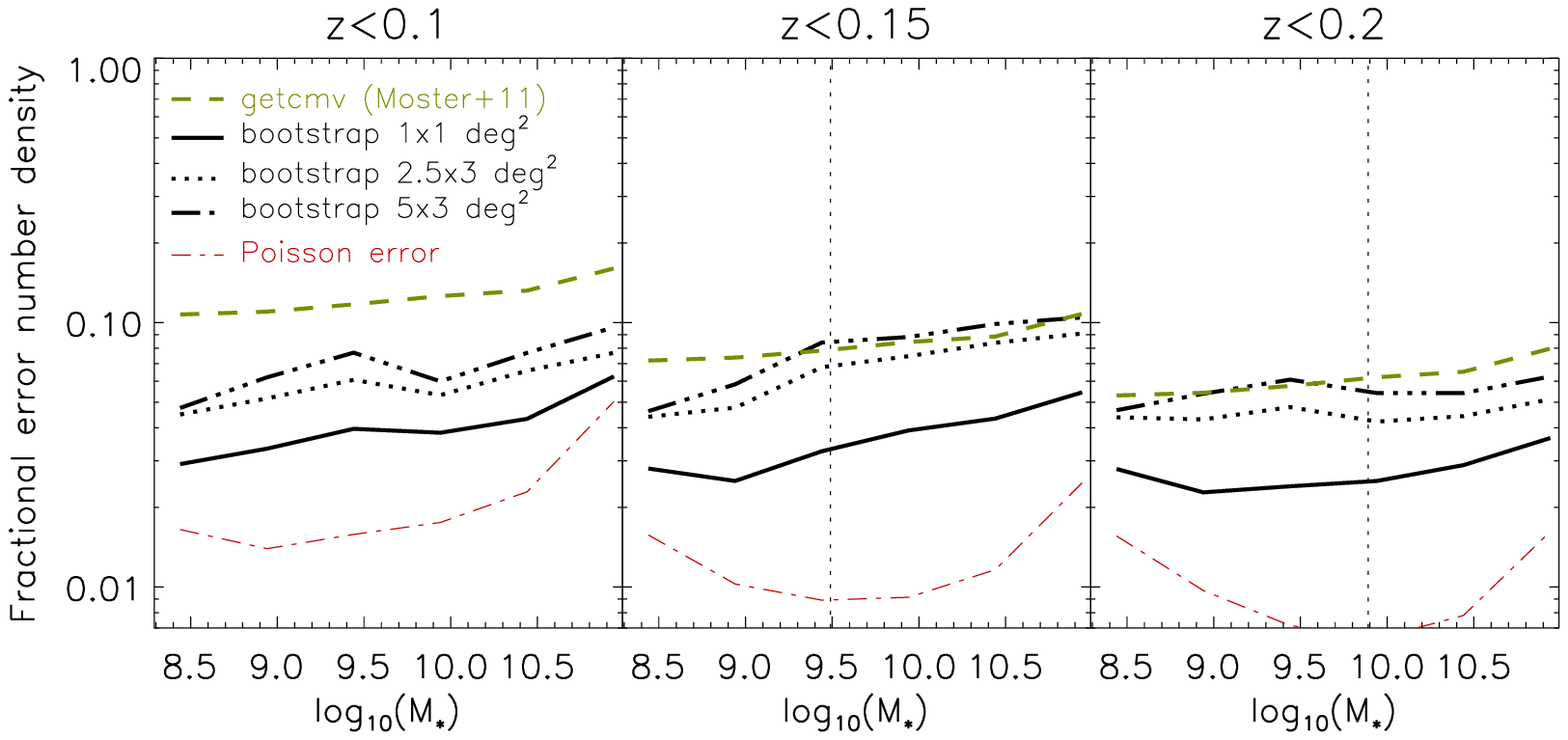}}
  \caption{Fractional error on stellar mass function measurements, using all GAMA galaxies below $z<0.1$, $z<0.15$ and $z<0.2$ for the left-hand, middle, and right-hand panel, respectively. Stellar masses are in units $\log_{10}(h^{-2}M_\odot)$. The black lines show the errors obtained from bootstrapping for three different patch sizes, as indicated in the plot. The red dotted line shows the Poisson contribution. The green dashed line shows the expected noise due to cosmic variance, predicted using the {\sc getcv} code from \citet{Moster11}. The vertical dotted lines show the approximate stellar mass completeness limits. This figure shows that cosmic variance is the dominant component in the stellar mass function errors, and that bootstrapping over too small volumes leads to underestimated error bars.}
  \label{plot_SMF_fracerr}
\end{figure*}
The errors on the stellar mass function are a combination of Poisson noise, cosmic variance and random errors in the stellar mass estimates. Since the latter is much smaller than the bin sizes of the stellar mass function, it should not affect the analysis. To estimate the combined error coming from Poisson noise and cosmic variance, we use a bootstrapping technique; we divide the GAMA catalogue into patches and randomly select subsamples to form new realisations of the data. We use $10\,000$ bootstrap realisations to ensure the results are converged. We experiment with different patch sizes and redshift cuts; the fractional errors are shown in Fig. \ref{plot_SMF_fracerr}, together with the Poisson noise contribution. This immediately reveals that the contribution of Poisson noise is subdominant compared to the contribution of cosmic variance. Secondly, it shows that the errors depend on the bootstrap patch size or volume; the larger the patch size, the larger the error. \\
\indent We compare our errors to the predictions from the {\sc getcv} code from \citet{Moster11}. Under the assumption that the galaxy bias is linear and independent of scale, the cosmic variance contribution to the stellar mass function is simply the product of the bias and the variance in the distribution of dark matter. We compute it for a patch size of 12$\times$5 deg$^2$ and divide the variance by 3, assuming that the three GAMA patches are independent. We show the predictions in Fig. \ref{plot_SMF_fracerr}. For $z<0.1$, the bootstrap errors are smaller than the prediction from \citet{Moster11}. For the higher redshifts, however, our errors using patches of 2.5$\times$3 deg$^2$ and 5$\times$3 deg$^2$ agree quite well with the predictions in the range where the stellar mass function is complete. This shows that the stellar mass function errors can be reliably determined via bootstrapping, as long as the volume of the bootstrapped samples is large enough. This may also explain why the jackknife errors on the stellar mass function in \citet{Coupon15} were a factor 2 smaller than the predicted errors, determined by combining the cosmic variance contribution from the {\sc getcv} code with the Poisson noise: their jackknifed volume was roughly a factor 2 (4) smaller than our 2.5$\times$3 deg$^2$ (5$\times$3 deg$^2$) patches at $z<0.15$. \\
\indent The covariance between the stellar mass function measurements has been ignored in most observational studies, even though \citet{Leauthaud11} have shown that it is important. The reason why the covariance is large, is simple: if the stellar mass function is a universal function, whose amplitude only differs due to local density variations, one would expect the measurements to be fully correlated. The main de-correlation mechanism is Poisson noise. Lower-level de-correlation happens due to the scale dependence and non-linearity of the bias. \\
\begin{figure*}
  \centering
  \resizebox{0.84\hsize}{!}{\includegraphics[angle=270]{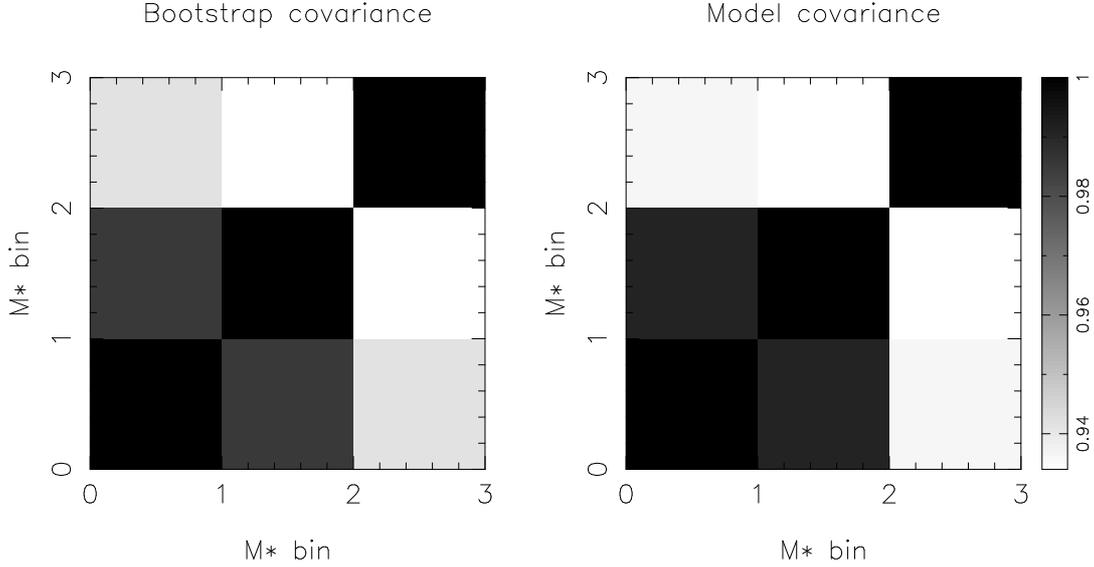}}
  \caption{Normalised correlation matrix of the stellar mass function, obtained using bootstrapping (left-hand panel), or modelled assuming that the measurements are fully correlated and the only de-correlation happens through Poisson noise (right-hand panel), as detailed in the text.}
  \label{plot_SMF_covar}
\end{figure*}
\indent We show the correlation matrix for three equally log-spaced stellar mass bins and $z<0.15$, determined by bootstrapping over 5$\times$3 deg$^2$ GAMA patches, in Fig. \ref{plot_SMF_covar}. The off-diagonals have values in the range 0.95-0.99, confirming that the stellar mass function measurements are highly correlated. To test whether the de-correlation is caused by Poisson noise, we have set the off-diagonals to one, added the Poisson noise contribution to the diagonals, and renormalised the covariance matrix. The resulting matrix is also shown in Fig. \ref{plot_SMF_covar}. The covariance matrix is very similar to the one obtained from bootstrapping, supporting the assumption that Poisson noise is mainly responsible for the de-correlation. \\
\indent Accounting for the covariance in the measurements is crucial. As is shown in \citet{Smith12}, the confidence contours of the parameters used to model the luminosity function (in that case, a Schechter function) change dramatically when the covariance is accounted for. We therefore determine the inverse of the sample covariance matrix, debiase that with a correction factor \citep{Kaufmann67,Hartlap07}, and use that to constrain the stellar mass function in the halo model.

\section{Sensitivity tests}\label{app_sens}

In this appendix, we detail on the sensitivity of our results on assumptions in the halo model.

\subsection{Stellar-to-halo mass relation of full sample \label{app_senslist}}
The tests we conducted are:
\begin{itemize}
\item Prior on $\beta_1$: We replaced the Gaussian prior on $\beta_1$ by a flat one in the range [0,15] and found that it did not impact the results. We only tested the impact of varying the prior on $\beta_1$, as the other parameters are constrained by the data.
\item Location of the knee of the satellite CSMF: In our fiducial set up, we fix the location of the knee in the CSMF of the satellites through $M^\rms_*=0.56 M^\rmc_*$. In principle, we could fit the location of the knee, although it cannot become arbitrarily large as that implies that the stellar mass of a satellite galaxy can be larger than that of the central. Although this could be avoided with the use of priors, we choose to avoid this issue altogether, fix the location of the knee and test for the sensitivity of this assumption. We replaced the location of the knee with $M^\rms_*=0.4 M^\rmc_*$ and $M^\rms_*=0.8 M^\rmc_*$, respectively. For $0.4 M^\rmc_*$, the fit slightly deteriorates (a minimum $\chi^2$ value of $\sim$88, compared to $\sim$80 for the fiducial run). The stellar-to-halo mass relation is slightly shallower at the high mass end (with a corresponding power law slope at the high-mass end of $\beta_2=0.30^{+0.03}_{-0.06}$), such that at a given stellar mass, galaxies reside in lower mass haloes. The shift is not significant. For $0.8 M^\rmc_*$, we obtain a best-fit $\chi^2$ value of $\sim$78. The stellar-to-halo mass relation steepens at the high mass end (with $\beta_2=0.21^{+0.04}_{-0.06}$), but not significantly so. 
\item Including a quadratic term in Eq. (\ref{eq:phi}): We included \mbox{$b_2 \times(\log_{10} M_{13})^2$} and also fit for $b_2$. As the best-fit $\chi^2$ value was virtually unchanged, the data do not require this term. At the high-mass end of the stellar-to-halo relation, the confidence intervals for the halo masses at a given stellar mass shift down by an insignificant amount of $\sim$0.5$\sigma$. 
\item Satellite distribution: We tested the assumption that the satellite distribution follows the dark matter. We adopted both a flatter and a steeper distribution, using $c_{\rm gal}=0.5c_{\rm dm}$ and $c_{\rm gal}=2.0c_{\rm dm}$, respectively, with $c_{\rm gal}$ the concentration of the satellite distribution and $c_{\rm dm}$ the one of the dark matter. The resulting stellar-to-halo mass relations were consistent with our nominal result. 
\item Subhalo mass: We assigned a zero subhalo mass to all satellites, $f_{\rm sub}=0$. Again, we found no significant changes compared to our nominal result.
\item Alternative stellar-to-halo mass relation: We adopted an alternative stellar-to-halo mass relation of the form:
\begin{equation}
M^\rmc_*(M_\rmh) = M_{*,0} {(M_\rmh/M_{\rmh,1})^{\beta_1} \over 
\left[1 + (M_\rmh/M_{\rmh,1})^{\beta_3} \right]^{(\beta_1-\beta_2)/\beta_3}}\,,
\label{eq_msmh2}
\end{equation}
and additionally fit for $\beta_3$. We adopted a Gaussian prior with zero mean and a width of 5. The best-fit $\chi^2$ takes a value of $\sim$75 (for 72 d.o.f.). We find $\beta_3 =3.32^{+4.58}_{-1.90}$, which is consistent with $\beta_3=1$, our fiducial set-up. The stellar-to-halo mass relation is slightly shallower at the high-mass end, but the 1$\sigma$ contours just overlap with our fiducial result. Adopting a stellar-to-halo mass relation of the form $M^\rmc_*(M_\rmh) = M_{*,0} (M_\rmh/M_{\rmh,1})^{\beta_1}$ leads to a best-fit $\chi^2$  of $\sim$220, hence such a model is strongly disfavoured by the data.

\end{itemize}

\subsection{Satellite fraction of full sample}
\indent We tested the impact of the list of halo model assumptions from Appendix \ref{app_senslist} on the recovered satellite fraction. Adopting $M^\rms_*=0.4 M^\rmc_*$ as the knee of the CSMF, the satellite fraction decreased on all scales by 0.02-0.04. Adopting Eq. (\ref{eq_msmh2}) as the stellar-to-halo mass relation resulted in larger satellite fractions, most noticeably at the high mass end, where the satellite fraction reached $0.09_{-0.05}^{+0.06}$. Changing the other assumptions led to smaller changes (of the order of a few per cent). \\

\subsection{Satellites in rich groups}
\indent We investigated how our results for satellites in `rich' groups changed when we varied the list of model assumptions from Appendix \ref{app_senslist}. Fixing the location of the knee of the satellite CSMF to \mbox{$M^\rms_*=0.4 M^\rmc_*$} significantly degraded the fit, with a minimum reduced $\chi^2$ value of 1.3 (compared to 0.9). Adopting $M^\rms_*=0.8 M^\rmc_*$ instead degraded the fit a little bit, resulting in $\chi_{\rm red}^2=1.1$. The confidence intervals of the satellite CSMF parameters shifted by up to 2.5$\sigma$ (compared to the constraints in Table \ref{tab_res}). The constraints on $f_{\rm conc}$ and $f_{\rm sub}$ did not change significantly. \\ 
\indent Changing the concentration of the satellite distribution did not affect the fit. The only fit parameter that was affected is $f_{\rm conc}$. For the $c_{\rm gal}=0.5c_{\rm dm}$ run, we obtained $f_{\rm conc}=1.35^{+0.25}_{-0.23}$, while for the $c_{\rm gal}=2c_{\rm dm}$ run, we found $f_{\rm conc}=0.67^{+0.13}_{-0.11}$. This shows that $f_{\rm conc}$ is partly constrained through $c_{\rm gal}$, the distribution of satellites, and that the concentration of the satellite distribution in `rich' groups is close to the concentration of the dark matter of the haloes that host them. \\
\indent When we included $b_2 \times(\log_{10} M_{13})^2$ in Eq. (\ref{eq:phi}) and fit for $b_2$, the models provided an equally good fit to the data, with $\chi_{\rm red}^2=0.9$, and we obtained $b_2=-0.41^{+0.26}_{-0.33}$. The constraints on $f_{\rm conc}$ and $f_{\rm sub}$ remained consistent with our nominal results. \\
\indent Enforcing a zero subhalo mass led to poor fits, where the model shear signal significantly underestimated the lensing data at small scales for the high stellar mass bins.

\subsection{Satellite fraction test of group galaxies}
We tested how the list of assumption in Appendix \ref{app_senslist} affected the halo model runs on centrals/satellites in `rich' groups in which we fitted for the satellite fraction. For the centrals, the satellite fraction remained consistent with zero, but the confidence intervals changed at \mbox{$M_* > 2\times10^{10} h^{-2}M_\odot$}: fixing the knee of the satellite CSMF to $M^\rms_*=0.4 M^\rmc_*$, the satellite fraction is constrained to $<0.01$ for the three highest stellar mass bins; when we adopted $c_{\rm gal}=0.5\times c_{\rm dm}$ the uncertainties increased instead and constrained to satellite fraction to $<0.15$. Changing the other assumptions led to smaller variations. For the satellites, the most extreme constraints came from the run where we enforced a zero subhalo mass fraction, where the satellite fraction of the highest stellar mass bin was $0.03^{+0.08}_{-0.03}$; when we fixed the knee of the satellite CSMF to $M^\rms_*=0.8 M^\rmc_*$, we obtained $0.15^{+0.32}_{-0.12}$, the other extreme.

\end{appendix}

\clearpage 

\end{document}